\documentclass[useAMS,usenatbib]{mn2e}
\usepackage{graphicx}
\usepackage{amsmath}
\usepackage{amssymb}

\title[{Chemical and structural analysis of the LMC}]{Chemical and structural 
analysis of the Large Magellanic Cloud using the fundamental mode RR Lyrae 
stars}
\author[Deb \& Singh]{Sukanta Deb\thanks{E-mail: 
sukantodeb@gmail.com}, Harinder P. Singh\\
Department of Physics \& Astrophysics, University of Delhi,
Delhi 110007, India\\
}
\begin{document}

\date{Received on ; Accepted on }

\pagerange{\pageref{firstpage}--\pageref{lastpage}} \pubyear{2013}

\maketitle

\label{firstpage}

\begin{abstract}
We present a careful and detailed light curve analysis of publicly available 
$I$-band data on fundamental mode RR Lyrae (RRab) stars of the Large 
Magellanic Cloud (LMC) obtained by the Optical Gravitational Lensing 
Experiment (OGLE) phase-III project. Using the Fourier parameters of 
$13,095$ RRab stars, metallicities and absolute magnitudes of individual stars are obtained. The 
representation of stars on the $P-\phi_{31}^{V}$ plane shows the existence of 
three significant metallicity groups with mean metallicities as 
$-1.20 \pm 0.12$ dex, $-1.57 \pm 0.10$ dex and $-1.89 \pm 0.09$ dex. 
The corresponding absolute magnitudes of these three 
groups are obtained as $0.70\pm 0.08$ mag, $0.59 \pm 0.06$ mag and 
$0.49 \pm 0.08$ mag, respectively. Distribution of these three groups as a 
function of vertical $|z|$ distance indicates that the formation of the LMC 
disk predates the formation of the inner halo. Issue of the existence of a 
metallicity gradient as a function of galactocentric distances has also been 
addressed.

Approximating the structure of the LMC disk as a triaxial ellipsoid, the inclination angle ($i$) relative to the plane 
of the sky and the position angle of the line of nodes ($\theta_{lon}$) were estimated 
as $24^{\circ}.20$ and $176^{\circ}.01$, respectively. The axes ratios and the eccentricity were 
also determined using the principal axes transformation method. 
\end{abstract}

\section{Introduction}
The presence of old stellar populations like the pulsating  RR Lyrae (RRL) 
stars, identified and characterized in large numbers in the LMC  by recent 
automated surveys such as OGLE, serve as an invaluable tool to unlock the 
secret of the galaxy. The core helium burning pulsating RRL stars are
excellent standard candles to estimate the Galactic and extragalactic 
distances and obey well defined luminosity-metallicity and 
period-luminosity-metallicity relations in optical and near infrared 
photometric bands \citep{butler03,catelan04,sollima06,catelan08,klein11}. In 
the $V$-band, the brightness of a RRL star is nearly standard with a slight 
metallicity dependence \citep{klein11}. 
They can be easily identified by their distinctive light 
curves. The presence of RRL stars in globular clusters facilitates the estimation 
of their ages and, hence, helps constrain the lower bound on the age of the 
Universe \citep{clementini10,de11,majaess12}. The RRL stars
are being used to determine the cosmological distance scale, 
the interstellar extinction along the line of sight  and to create a 
three dimensional map of different galaxies \citep{as_smc,haschke12}.             

The LMC serves as an important target in the calibration of cosmic distance 
scale because of its proximity and low inclination angle. This galaxy hosts a 
statistically large sample of `standard candles' which includes RRL, Cepheids 
and Red Clump giant stars. The accurate distance determination to the LMC 
plays a significant role in constraining the value of the Hubble constant 
$H_{0}$ \citep{schaefer08,de11,riess11}. The distance measurement to the 
LMC has revolutionized our understanding of the distance scale of the Universe 
and supported the evidence for the expansion of the Universe \citep{chall12}. 
The unique location of the LMC along with the abundance of various types of 
`standard candles' allows us to compare and calibrate a large sample of 
distance indicators, which in turn can be utilized for more distant objects 
\citep{de11}.

Generally, the distances can be measured more accurately from the binary 
star light curve modelling using the combined photometric and spectroscopic 
data \citep{vilardell10,ds11,chall12,nature13}. Recently, using the 
surface-brightness/color relation of the components of eight binary systems in 
the LMC, the angular sizes of the components were combined with their linear 
dimensions obtained from the light curve modeling to obtain the 
distance modulus of the LMC as 
$(m-M)_{0}=18.493\pm0.008(\text{statistical})\pm0.047(\text{systematic})$ mag
\citep{nature13}. This corresponds to a distance of
$49.97\pm0.19(\text{statistical}) \pm1.11(\text{systematic})$ kpc,  
accurate to $2\%$. Most of the systems were located near the center of the 
LMC and situated along the line of nodes \citep{nature13}.              
           
The LMC is a late type disk galaxy seen nearly face-on with enormous 
amount of gas, dust and inhabits sites for active star formation. There is a 
spectacular evidence of onging interactions with both the Milky Way and the 
Small Magellanic Cloud \citep{lin95,westerlund97,as10}. The depth of the LMC is a few kpc 
along the line of sight. The geometry and orientation of the disk galaxy such 
as LMC  has been the subject of numerous studies through the use of different 
tracers. The issue of the LMC geometry has major implications. Until
the last decade, our understanding of the LMC was shallow due to 
the small number of tracers and availability of the data with limited spatial 
coverage \citep{vand101,wein01,as03,niko04,as09,cioni09,as13}. The LMC 
has been the best astrophysical laboratory for studies of various stellar 
populations, interstellar matter, star-formation processes and the galactic 
structure at large \citep{fukui10,glatt10}.    

The OGLE database has become a valuable resource for studying the properties 
of variable stars as it contains a pool of light curve data of a variety of 
variables present in our Galaxy as well as the  nearby satellite galaxies, 
viz., the LMC and the SMC (Small Magellanic Cloud). The present paper attempts to use the wealth of RRL 
data of LMC from the OGLE-III catalog in order to unravel the various chemical 
and structural properties of this galaxy. A proper and systematic light curve 
analysis of the LMC RRab stars has been carried out in order to comprehend the 
metallicity and distance distribution of these tracers. These parameters along 
with the others have been utilized in deciphering the structure of the LMC. 
The independent analysis based on the RRab stars done in this paper will serve 
as a source to compare and correlate with the determinations of 
structural parameters of the LMC, such as inclination ($i$) and position angle 
of the line of nodes ($\theta_{lon}$), derived from other studies using 
different tracers and methodologies.           

In order to make the OGLE data more useful, we have cleaned the phased light 
curves using $2\sigma$ clipping. The clean phased light curves were then 
Fourier decomposed in order to determine the various parameters and their 
associated errors. This paper will, thus, serve as a supplementary and 
advanced version of the OGLE catalog of the LMC RRab stars, in which all the 
useful Fourier parameters are provided. 
      
The Fourier parameters determined from the light curves of RRab stars are 
useful for distance determinations and metal abundance of the galaxy in which 
they are present. The past research has proved that the shape of a RRL 
light curve can be described in terms of the Fourier parameters that can 
be linked with the physical parameters such as mass ($M$), radius ($R$), and 
luminosity ($L$) of the star. They can also be linked to the other 
atmospheric parameters, such as metallicity ($[Fe/H]$), effective temperature 
$(T_{eff})$ and surface gravity ($\log{g}$) of the RRL stars 
\citep{jurc96,kova96,jurc98,kova98,kova01}. Nevertheless, it has also been 
shown by \citet{cacc05} that the intrinsic colors derived from Fourier coefficients
show discrepancies with the observed ones, and hence the resulting temperatures 
and temperature-related parameters are unreliable.     

The paper is organized as follows. In section~\ref{data}, we describe the 
sample of RRab stars in the OGLE-III LMC database \citep[hereafter, SZ09]
{sosz09}. The Fourier decomposition method as applied to the RRab light curves 
in the $I$-band and the sample selection criteria are described in 
section~\ref{fd}. Section~\ref{pparameters} describes the determination of metallicities and absolute magnitudes of the RRab stars using the 
calibration relations between $I$ and  $V$-bands and then using the resulting 
$V$-band parameters in the empirical relations from the literature. Distance 
determination to individual RRab stars of the LMC using the 
intensity weighted mean magnitudes, the absolute magnitudes and the 
interstellar extinction has also been 
discussed. 
In section~\ref{structure}, we describe the determination of the structural 
parameters of the LMC by plane fitting procedure and principal axes 
transformation method using the moment of inertia tensor constructed from the 
projected three dimensional Cartesian coordinates of RRab stars. Dependence of 
the geometrical parameters on the choice of the LMC center is discussed. 
Comparison with other studies in the literature is provided. 
Section~\ref{metgrad} discusses the issue of metallicity gradient in the LMC 
as a function of galactocentric distance ($R_{GC}$) using the metallicity 
values obtained from four different empirical relations available in the 
literature. Finally, in section~\ref{summary}, the summary and the conclusions 
of the present study are laid down. 
\section{The Data}
\label{data}
\begin{figure}
\includegraphics[width=0.5\textwidth,keepaspectratio]{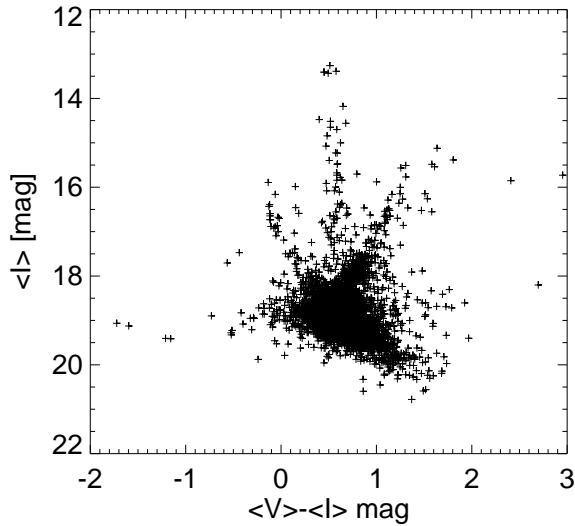}
\caption{Color-magnitude diagram of the $17,337$ RRab stars which have both $I$
and $V$-band light curves available in the OGLE-III catalog.}
\label{color_mag}
\end{figure}
\begin{figure}
\includegraphics[width=0.5\textwidth,keepaspectratio]{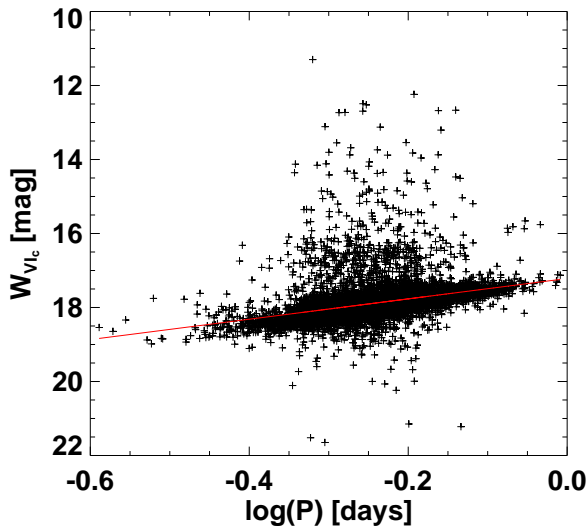}
\caption{Wesenheit index ($W_{VI_{c}}$) as a function of pulsation period 
($P$) of the $17,337$ LMC RRab stars which have both $I$ and $V$-band light 
curves available in the OGLE-III catalog. Solid line represents the 
fit to the observed data points from the empirical relation of \citet{sosz03}
as given by Eqn.~\ref{wes}.}
\label{wesenheit}
\end{figure}
We selected RRab stars from OGLE-III catalog that consists of  $8$-year 
archival data identified and characterized by the Fourier coefficients 
of the light curves (SZ09).  The catalog contains $17,693$ RRab stars having a 
mean period of $<P_{ab}> = 0.576$ days. The OGLE field in the LMC covers nearly $
40$ $\deg^{2}$. Most of the observations were carried out using the Cousins 
$I$-band filter with exposure time of $180$s having an average of $400$ 
photometric observations. The catalog also contains $V$-band light curves 
of $17,337$ stars having an average of $30$ data points per light curve. The 
color ($<V>-<I>$)-magnitude ($<I>$) diagram for $17,337$ stars is shown in 
Fig.~\ref{color_mag}, where $<V>$ and $<I>$ are the intensity weighted mean 
magnitudes in the $V$ and $I$-band, respectively and are obtained from the 
Fourier fitted light curves. Fig.~\ref{wesenheit} shows the plot of reddening 
free Wesenheit index ($W_{VI_{c}}$) as a function of pulsation period ($P$) of 
the $17,337$ RRab stars, where $W_{VI_{c}}$ is defined as 
$W_{VI_{c}} = <I>-1.55(<V>-<I>)$ \citep{sosz03}. Also overplotted on the data points is the 
fit from the empirical relation \citep{sosz03}
\begin{equation} 
\label{wes}
W_{VI_{c}} = (-2.75\pm0.04)\log{P} +(17.217\pm0.008).
\end{equation}
The empirical slope of the function is consistent with that predicted by 
models \citep{cris04}.

For the analysis of light curves of RRab stars, we have selected the 
$I$-band data as they contain relatively larger number of data points per 
light curve as compared to corresponding $V$-band light curves. The OGLE-III 
light curve data of the LMC central region are further supplemented by the 
OGLE II data collected between $1997$ and $2000$ using the same telescope set 
up. The catalog contains intensity $I$ and $V$-band mean magnitudes, 
periods and their uncertainties in days, epochs of maximum 
light, peak-to-peak $I$-band magnitudes and the Fourier parameters $R_{21}, 
\Phi_{21}, R_{31}$  and $\Phi_{31}$ derived from the $I$-band light curves. 
\begin{figure*}
\begin{center}
\includegraphics[width=1.0\textwidth,keepaspectratio]{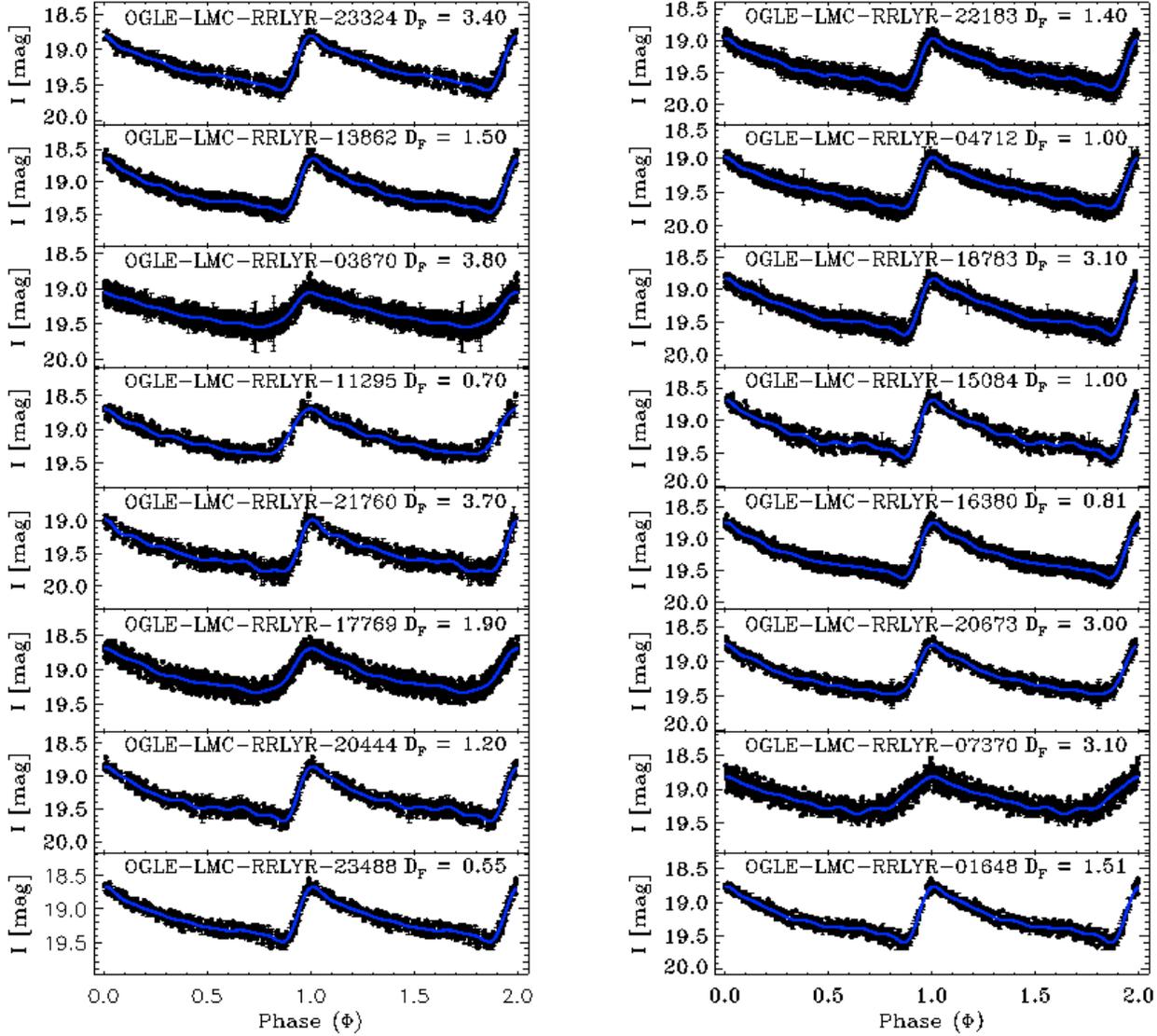}
\caption{Fourier fitted RRab light curves in the $I$-band from the 
catalog. The quantity $D_{F}$ represents the deviation parameter as defined 
by \citet{jurc96} and is discussed in section~\ref{iron}.}
\label{fitv}
\end{center}
\end{figure*}
\section{Fourier decomposition method and sample selection}
\label{fd}
The Fourier decomposition method described below was used to obtain various light curve 
parameters of the LMC RRab stars. The light curves were fitted with a Fourier 
cosine series of the form
\begin{equation}
\label{fdtech}
m(t) = A_{0}+\sum_{i=1}^{N} A_{i} \cos(i\omega (t-t_{0}) +  \phi_{ i}),
\end{equation} where $m(t)$ is the observed magnitude, $A_{0}$ is the mean magnitude, $\omega$=2$\pi/P$ is the angular frequency, $P$ is the period of the star
 in days and $t$ is the time of observation. The epoch of maximum light 
$t_{0}$ is used to obtain a phased light curve which has maximum light at 
phase zero. $A_{i}$'s and $\phi_{i}$'s are the $i$th order Fourier coefficients 
and $N$ is the order of the fit. Eqn.~\ref{fdtech} has $2N+1$ unknown 
parameters. To solve for these parameters, we require at least the same number 
of data points. The light curves were phased using
\begin{displaymath}
\Phi =\frac{\left( t-t_{0}\right) }{P}-Int\left( \frac{\left(
t-t_{0}\right) }{P}\right).
\end{displaymath}  
The pulsation periods ($P$) and the $t_{0}$ values are taken from the OGLE 
catalog. An automated code was developed  using the MPFIT package in IDL  
to obtain various Fourier parameters on the 
right hand side of Eqn.~\ref{fdtech} \citep{ds10}. A seventh order 
Fourier fit was employed to model the RRab $I$-band light curves. 
The Fourier fitted $I$-band light curves of a sample of RRab stars are shown
in Fig.~\ref{fitv}. The phase
differences, $\phi_{i1} = \phi_{i}-i\phi_{1}$ and amplitude ratios,
$R_{i1} = (A_{i}/A_{1})$, $i > 1$ were evaluated and standard errors
were determined using the formulae of \citet{ds10}.

\begin{figure}
\includegraphics[width=0.5\textwidth,keepaspectratio]{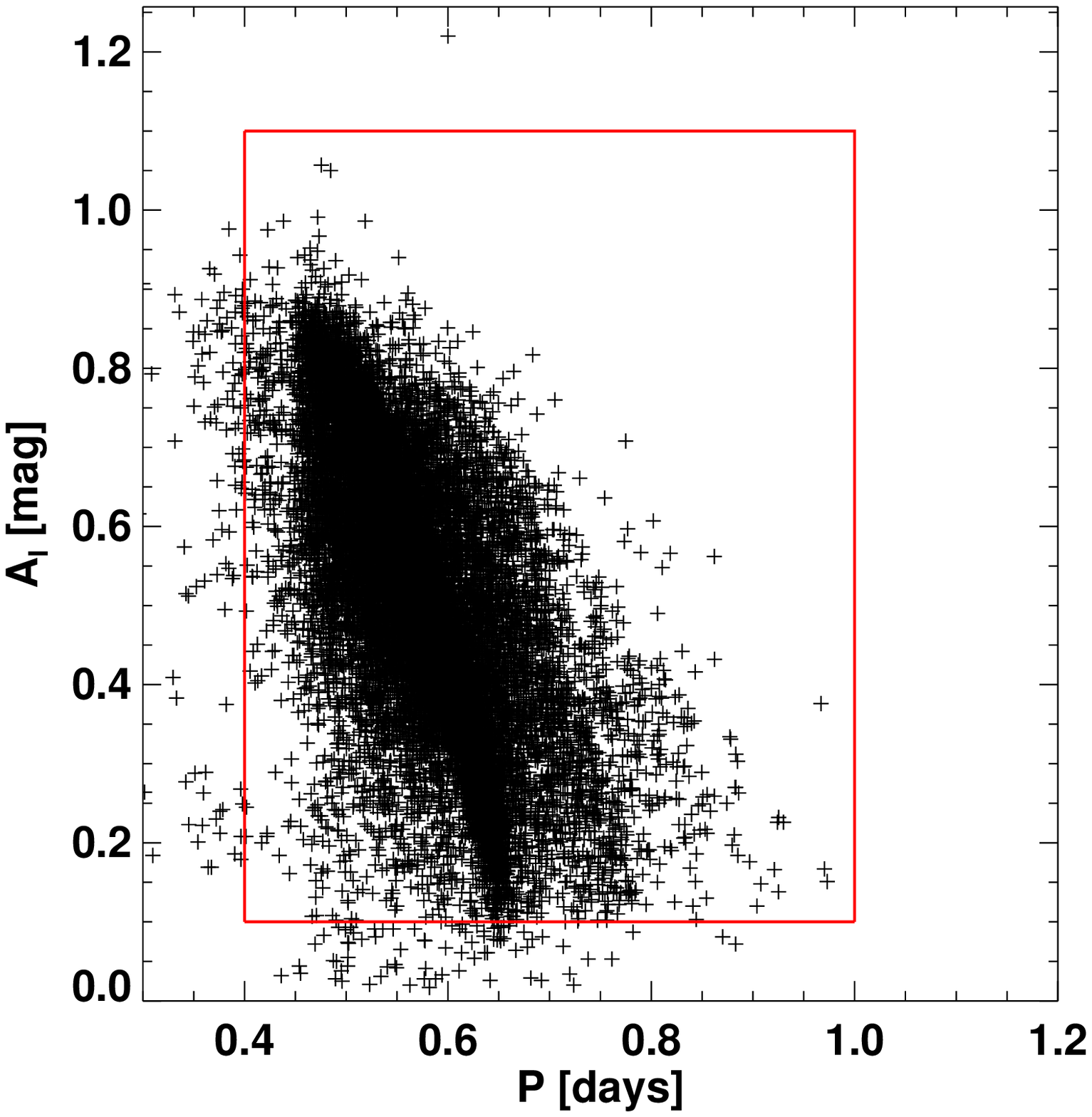}
\caption{Period-Amplitude diagram for $17,693$ stars in the OGLE database. The retained $17,092$ RRab stars filtered through the 
selection criteria on period, amplitude and mean magnitude lie in the inner rectangle.}
\label{period_amp}
\end{figure}

\begin{figure}
\begin{center}
\includegraphics[width=0.5\textwidth,keepaspectratio]{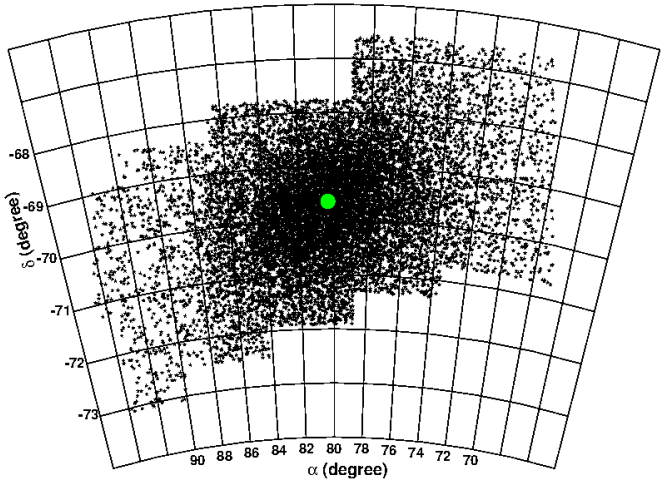}
\caption{Equatorial coordinates $(\alpha,\delta)$ of the selected sample of 
$13,095$ RRab stars of the LMC. The filled circle denotes the position 
of the centroid $(\alpha_{0},\delta_{0}) = (80^{\circ}.35, -69^{\circ}.65)$.} 
\label{curvilinear}
\end{center}
\end{figure}

In order to have a clean sample of RRab stars for the present analysis, we 
have used the selection criteria based on OGLE-determined periods, the mean 
magnitude ($A_{0}$) and the peak-to-peak $I$-band amplitude ($A_{I}$) determined
from the Fourier analysis of the cleaned phased light curves. We chose RRab 
stars with period $P \ge 0.4$ days, mean magnitude $A_{0} \ge 17$ mag and 
amplitude $0.1 \le A_{I} \le 1.2$ mag. These selection criteria were applied 
in order to exclude the RRc stars and the high amplitude $\delta$ Scuti stars 
that may have been misclassified as RRab stars, thus ensuring a reliable 
sample of RRab stars for the present analysis. The application of these 
selection criteria reduces the number of RRab stars in the $I$-band light 
curves to $17,092$. The period-amplitude diagram for the complete and the 
selected sample of $17,092$ RRab stars is shown in Fig.~\ref{period_amp}. In 
order to select the `normal-looking' RRab stars out of $17, 092$ stars, the 
application of the compatibility test of \citet{jurc96} further reduces the 
number of stars to $13,095$. These $13,095$ stars were retained for the 
physical parameter estimation and determination of the structure of the LMC in 
the present study. The equatorial coordinates ($\alpha, \delta$) of the selected
sample of $13,095$ RRab stars are shown in Fig.~\ref{curvilinear}. The centroid of the final sample is 
$(\alpha_{0},\delta_{0}) = (80^{\circ}.35, -69^{\circ}.65)$ and is shown as a 
filled circle in the figure. The Fourier parameters obtained from the 
Fourier decomposition method are listed in Tables~{\ref{four1}} and 
~{\ref{four2}}, respectively.


\begin{figure}
\includegraphics[width=0.5\textwidth,keepaspectratio]{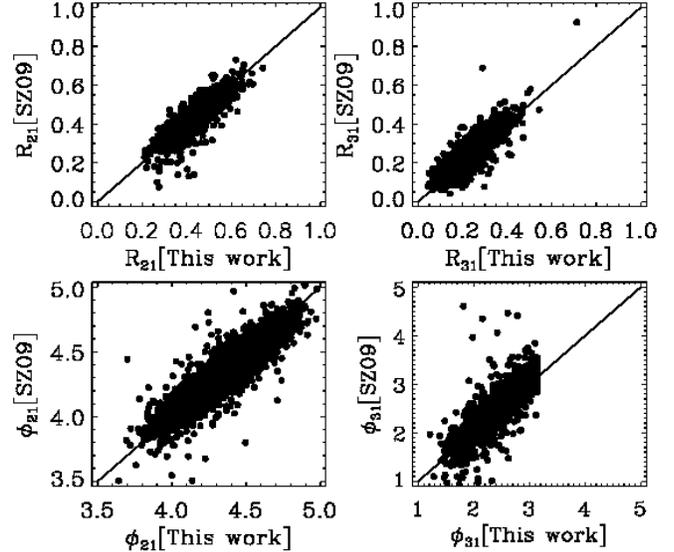}
\caption{A comparison of the Fourier parameters $R_{21}$, $R_{31}$,
$\phi_{21}$, $\phi_{21}$ of $13,095$ stars from \citet{sosz09} and the present
study.}
\label{pap_lit}
\end{figure}
\begin{table*}
\centering
\caption{A sample of the Fourier parameters of $13,095$  OGLE RRab stars 
with increasing period used for the analysis obtained from the Fourier 
cosine decomposition of the $I$-band light curves as given by 
Eqn.~\ref{fdtech}. The full table is available as supplementary material in 
the online version of this paper.}
\label{four1}
\scalebox{0.8}{
\begin{tabular}{ccccccccccc}
\\
\hline
\hline
OGLE ID & P[days] & $A_{0}$& $A_{1}$ & $A_{2}$ & $A_{3}$ & $A_{4}$ & $\phi_{1}$ &$\phi_{2}$ & $\phi_{3}$ & $\phi_{4}$   \\
&&$\sigma A_{0}$& $\sigma A_{1}$ & $\sigma A_{2}$ & $\sigma A_{3}$ & $\sigma A_{4}$ & $\sigma \phi_{1}$ &$\sigma \phi_{2}$ & $\sigma \phi_{3}$ & $\sigma \phi_{4}$\\\hline
OGLE-LMC-RRLYR-23488& 0.4003274&  19.16019&   0.25560&   0.13633&   0.09880&   0.06321&   2.27497&   2.40845&   2.77793&   3.21414\\
&&   0.00427&   0.00529&   0.00586&   0.00584&   0.00596&   0.02633&   0.04545&   0.06294&   0.09666\\
OGLE-LMC-RRLYR-20444& 0.4005525&  19.35242&   0.25310&   0.15458&   0.10378&   0.05028&   2.33045&   2.35537&   2.79613&   3.13930\\
&&   0.00509&   0.00641&   0.00719&   0.00721&   0.00719&   0.03139&   0.04714&   0.06972&   0.14416\\
OGLE-LMC-RRLYR-17769& 0.4009905&  19.07511&   0.24234&   0.11507&   0.04785&   0.01949&   2.49946&   2.78721&   3.39620&   3.82907\\
&&   0.00173&   0.00229&   0.00239&   0.00244&   0.00242&   0.01071&   0.02173&   0.05121&   0.12683\\
OGLE-LMC-RRLYR-21760& 0.4011895&  19.48245&   0.25007&   0.14082&   0.09722&   0.05027&   2.25690&   2.41177&   2.97533&   3.15050\\
&&   0.00548&   0.00713&   0.00765&   0.00764&   0.00768&   0.03306&   0.05525&   0.08149&   0.15664\\
OGLE-LMC-RRLYR-11295& 0.4017449&  19.11341&   0.25864&   0.11954&   0.05925&   0.03505&   2.50487&   2.94624&   3.36518&   3.90536\\
&&   0.00399&   0.00492&   0.00509&   0.00528&   0.00547&   0.02415&   0.05108&   0.09987&   0.16220\\
OGLE-LMC-RRLYR-03670& 0.4017562&  19.33933&   0.19194&   0.07938&   0.03781&   0.02109&   2.41474&   2.83343&   3.41382&   3.72526\\
&&   0.00299&   0.00408&   0.00417&   0.00420&   0.00421&   0.02274&   0.05401&   0.11244&   0.20248\\
OGLE-LMC-RRLYR-13862& 0.4018669&  19.15250&   0.26409&   0.14338&   0.09822&   0.05805&   2.31965&   2.37446&   2.84023&   3.06405\\
&&   0.00226&   0.00295&   0.00315&   0.00315&   0.00313&   0.01298&   0.02263&   0.03305&   0.05609\\
OGLE-LMC-RRLYR-23324& 0.4019162&  19.25332&   0.24472&   0.13560&   0.08229&   0.05855&   2.28630&   2.50185&   3.08952&   3.35738\\
&&   0.00458&   0.00642&   0.00625&   0.00645&   0.00636&   0.02656&   0.04917&   0.07859&   0.11124\\
OGLE-LMC-RRLYR-01648& 0.4020618&  19.24330&   0.25264&   0.15037&   0.09507&   0.05554&   2.25824&   2.30472&   2.71777&   3.19910\\
&&   0.00407&   0.00512&   0.00566&   0.00571&   0.00550&   0.02521&   0.03935&   0.06000&   0.10625\\
OGLE-LMC-RRLYR-07370& 0.4025514&  19.13236&   0.21665&   0.06850&   0.03027&   0.01456&   2.65053&   3.18102&   3.46347&   3.24620\\
&&   0.00241&   0.00327&   0.00341&   0.00346&   0.00343&   0.01641&   0.04971&   0.11048&   0.23261\\
\hline
\hline
\end{tabular}
}
\end{table*}

\begin{table*}
\centering
\caption{A sample of the Fourier parameters of $13,095$  OGLE RRab stars 
with increasing  period used for the analysis obtained from the Fourier 
cosine decomposition of the $I$-band light curves as given by 
Eqn.~\ref{fdtech}. The full table is available as supplementary material in 
the online version of this paper.}
\label{four2}
\scalebox{0.8}{
\begin{tabular}{ccccccccccccc}
\\
\hline
\hline
OGLE ID & $N_{obs}$& $\chi_{\nu}^{2}$& $\sigma_{fit}$ & $A_{I}$ & $R_{21}$ & $R_{31}$ & $R_{41}$ &$\phi_{21}$ & $\phi_{31}$ & $\phi_{41}$   \\
&&&&$\sigma A_{I}$& $\sigma R_{21}$ & $\sigma R_{31}$ & $\sigma R_{41}$ & $\sigma \phi_{21}$ & $\sigma \phi_{31}$ &$\sigma \phi_{41}$ \\\hline
OGLE-LMC-RRLYR-23488&  319&   0.73841&   0.05952&   0.82006&   0.53339&   0.38655&   0.24731&   4.14170&   2.23621&   0.39746\\
&&&&   0.10851&   0.02544&   0.02420&   0.02388&   0.05252&   0.08206&   0.12483\\
OGLE-LMC-RRLYR-20444&  313&   0.67103&   0.06892&   0.83119&   0.61074&   0.41004&   0.19865&   3.97765&   2.08796&   0.10067\\
&&&&   0.12521&   0.03235&   0.03033&   0.02885&   0.05664&   0.09382&   0.17220\\
OGLE-LMC-RRLYR-17769&  947&   2.73231&   0.08563&   0.64168&   0.47482&   0.19746&   0.08044&   4.07149&   2.18101&   0.11443\\
&&&&   0.09974&   0.01084&   0.01024&   0.01001&   0.02422&   0.05551&   0.13083\\
OGLE-LMC-RRLYR-21760&  308&   0.83825&   0.08641&   0.79528&   0.56310&   0.38876&   0.20104&   4.18116&   2.48782&   0.40610\\
&&&&   0.03947&   0.03457&   0.03251&   0.03124&   0.06438&   0.10494&   0.18540\\
OGLE-LMC-RRLYR-11295&  319&   1.48647&   0.07479&   0.67497&   0.46218&   0.22907&   0.13552&   4.21969&   2.13376&   0.16907\\
&&&&   0.04823&   0.02157&   0.02089&   0.02129&   0.05650&   0.11093&   0.17764\\
OGLE-LMC-RRLYR-03670&  889&   0.91742&   0.08414&   0.49276&   0.41359&   0.19701&   0.10987&   4.28713&   2.45277&   0.34946\\
&&&&   0.08728&   0.02343&   0.02230&   0.02205&   0.05860&   0.12129&   0.21366\\
OGLE-LMC-RRLYR-13862&  797&   1.23480&   0.06661&   0.82504&   0.54294&   0.37191&   0.21980&   4.01833&   2.16446&   0.06862\\
&&&&   0.05998&   0.01337&   0.01263&   0.01212&   0.02609&   0.04203&   0.06829\\
OGLE-LMC-RRLYR-23324&  309&   0.81679&   0.06638&   0.76751&   0.55410&   0.33626&   0.23924&   4.21244&   2.51381&   0.49538\\
&&&&   0.09538&   0.02940&   0.02778&   0.02675&   0.05588&   0.09486&   0.13684\\
OGLE-LMC-RRLYR-01648&  322&   0.74707&   0.05604&   0.83052&   0.59520&   0.37631&   0.21983&   4.07143&   2.22625&   0.44934\\
&&&&   0.06144&   0.02545&   0.02384&   0.02223&   0.04673&   0.07837&   0.13042\\
OGLE-LMC-RRLYR-07370&  635&   3.23408&   0.09796&   0.55133&   0.31618&   0.13971&   0.06720&   4.16315&   1.79507&   5.21046\\
&&&&   0.17864&   0.01646&   0.01613&   0.01585&   0.05235&   0.11525&   0.23776\\
\hline
\hline
\end{tabular}
}
\end{table*}
In order to make a comparison of the parameters obtained in the present study with those given in the OGLE-III catalog,  
scatter plots of the Fourier amplitudes $R_{21}, R_{31}$ and the phase parameters $\phi_{21}, \phi_{31}$ 
are shown in Fig.~\ref{pap_lit}.
\begin{figure}
\includegraphics[width=0.5\textwidth,keepaspectratio]{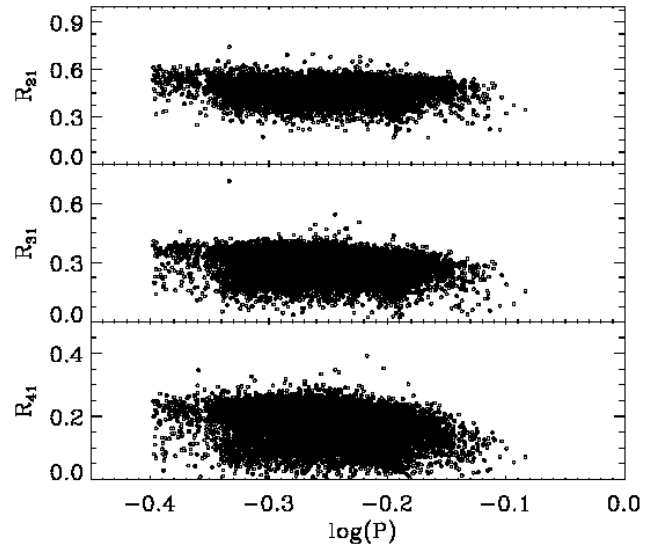}
\caption{Fourier amplitude ratios $R_{21}, R_{31}$ and $R_{41}$ as a function 
of $\log{(P)}$ for the $I$-band data.}
\label{r21_per}
\end{figure}
\begin{figure}
\includegraphics[width=0.5\textwidth,keepaspectratio]{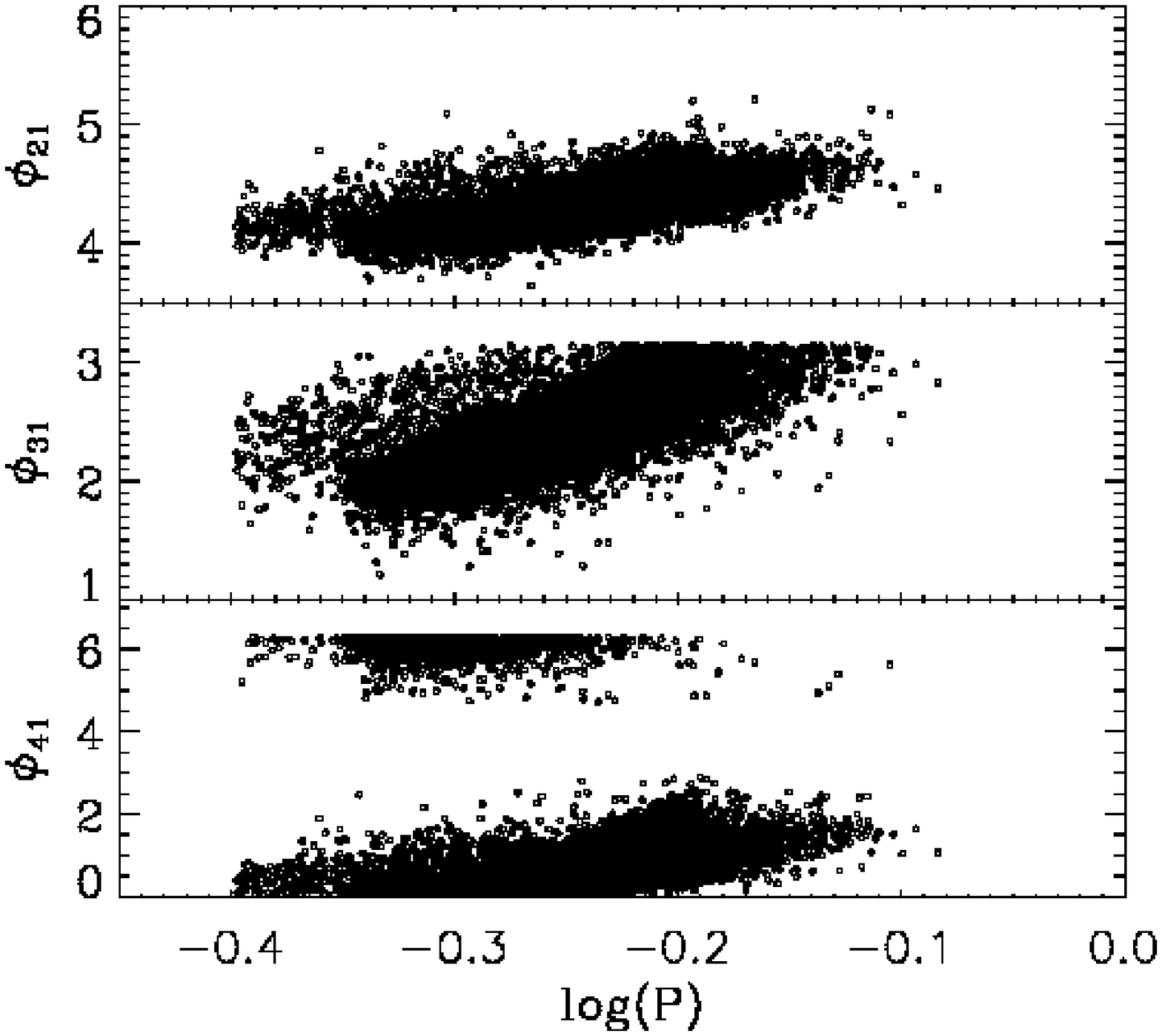}
\caption{Fourier phase parameters $\phi_{21}, \phi_{31}$ and $\phi_{41}$ as a 
function of $\log{(P)}$ for the $I$-band data.}
\label{phi21_per}
\end{figure}
For the $I$-band data under consideration, the Fourier amplitude ratios $R_{21}, R_{31}$ and $R_{41}$ as a function 
of $\log{(P)}$ are shown in Fig.~\ref{r21_per}.
The plot of the Fourier phase parameters  $\phi_{21}, \phi_{31}$ and 
$\phi_{41}$ as a function of $\log{(P)}$ is shown in Fig.~\ref{phi21_per}.       
The dispersion in the $\phi_{i1}$ values ($i=2,3,4$) goes on increasing as we 
go to higher orders. This is because the uncertainty in the determination of 
the Fourier phase parameters increases as one goes to higher orders. 
Fig.~\ref{phi21_per} illustrates that $\phi_{21}$ and $\phi_{31}$ increase 
with increasing period. On the contrary, there is no clear trend for 
$\phi_{41}$ which is accounted for by their large uncertainties.

\begin{figure}
\includegraphics[width=0.5\textwidth,keepaspectratio]{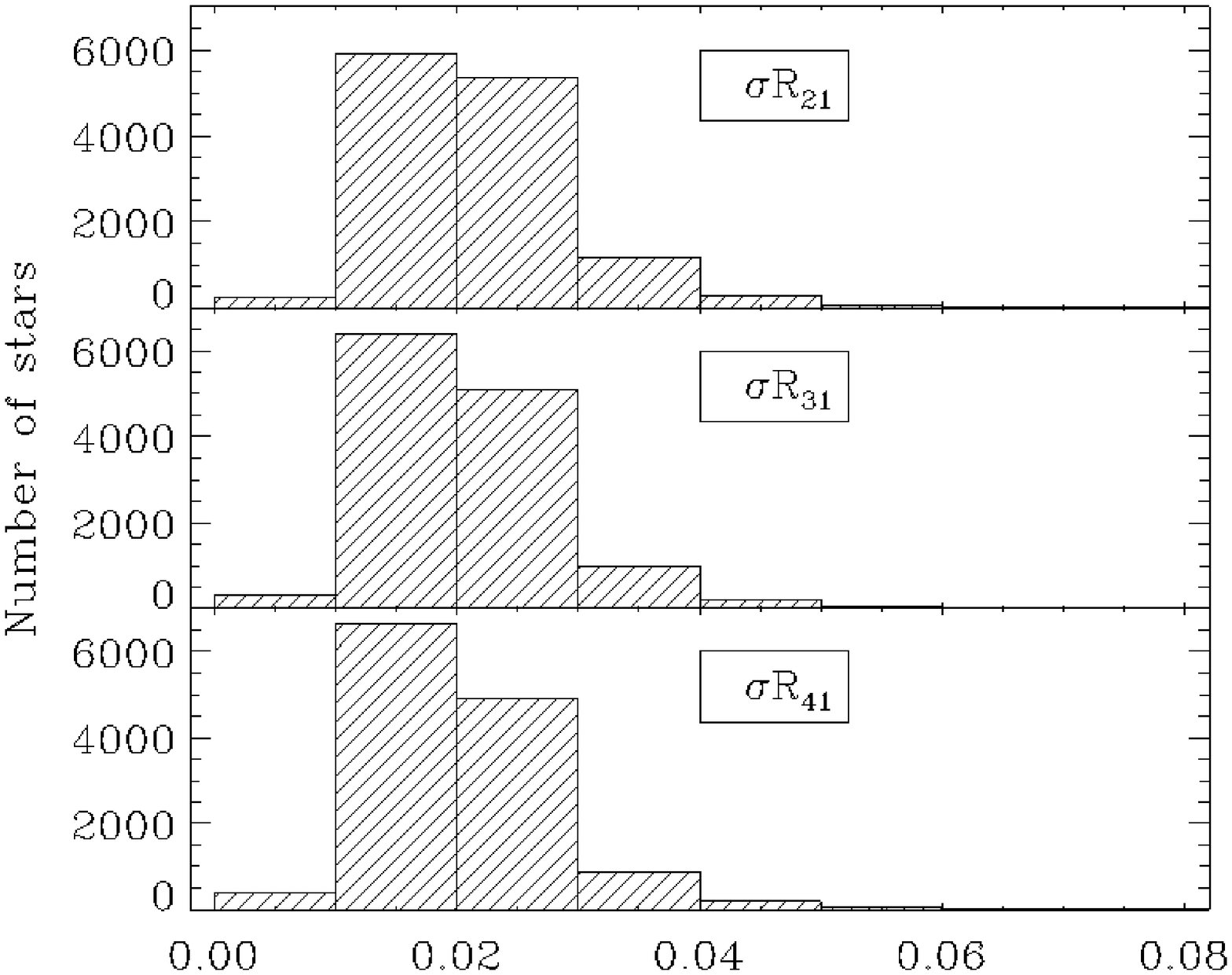}
\caption{Histogram plot of the distribution of  standard errors in $R_{21}, 
R_{31}$ and $R_{41}$ for the $I$-band data.}
\label{er21}
\end{figure}

\begin{figure}
\includegraphics[width=0.5\textwidth,keepaspectratio]{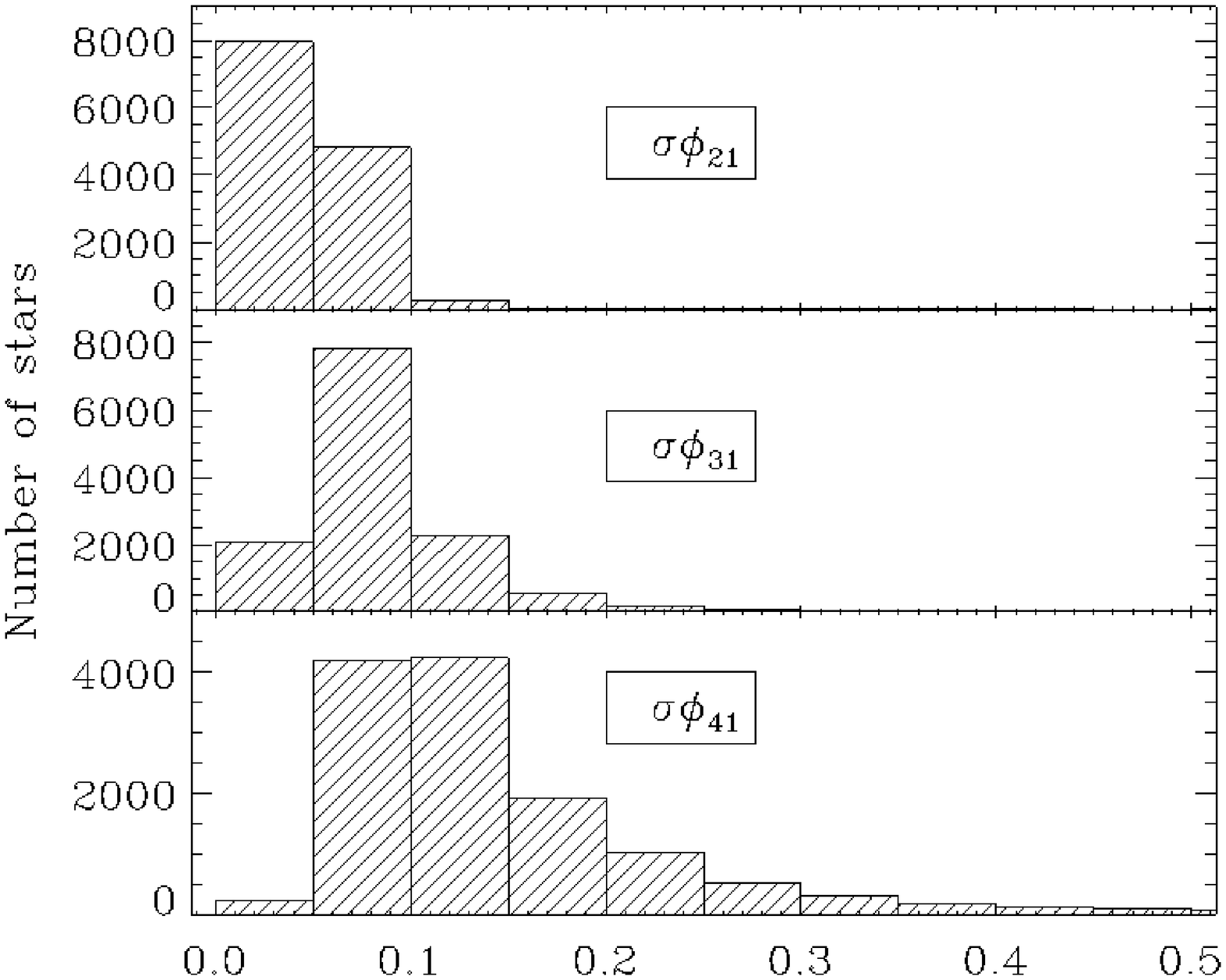}
\caption{Histogram plot of the distribution of standard errors in 
$\phi_{21}, \phi_{31}$ and $\phi_{41}$ for the $I$-band data.}
\label{ephi21}
\end{figure}

In Fig.~\ref{er21}, the estimated errors for $R_{21}, R_{31}$, and $R_{41}$ 
are plotted. The distribution of the errors are 
similar for all the three Fourier amplitudes. On the other hand, from Fig.~\ref{ephi21} we can see 
that the errors in $\phi_{41}$ are larger than those of $\phi_{21}$ and 
$\phi_{31}$, whereas, in turn the errors in $\phi_{31}$ are larger than that 
of $\phi_{21}$. This is expected as the amplitudes for the higher orders become 
smaller and smaller, it becomes increasingly difficult to derive their phases with high precision.

 \section{Determination of Physical Parameters}
\label{pparameters}
One of the major constraints in deriving the physical parameters of RRL stars from $I$-band data is that most of the empirical relations connecting the Fourier parameters and the 
physical parameters are derived from the $V$-band RRL light curves.
However, the huge RRab data obtained from various automated 
surveys and space missions are not limited to the Cousins $V$-band \citep{udal97,wozniak04}.  It is, therefore, quite natural to look for some inter-relations between these parameters obtained 
from  a considerable number of well-sampled light curves, both in the $V$ band as well as  any other photometric band like the $I$-band.
In the literature, there are several instances, where the Fourier 
parameters in different bands are scaled to the corresponding values in the 
standard $V$-band in order to make use of the useful empirical relations. For 
example, the Fourier phase parameter values $\phi_{31}$ for the Northern Sky 
Variability Survey (NSVS) photometric system were calibrated to the standard 
Johnson $V$ photometric system using the $55$ well observed RRab stars which had 
the compiled values of $\phi_{31}$ in the $V$-band \citep{karen06}. 
\citet{ds10} also derived interrelations between various parameters of the 
RRL light curves in the $V$ and $I$-band with highly accurate complementary 
light curves in the literature in order to derive the various physical 
parameters of the SMC RRL stars. \citet{nemec11} also used the $V$-band 
empirical relationships between the Fourier parameters and the physical 
parameters in order to derive the various properties of the RRab stars 
observed with Kepler, by noting that the two broad 
photometric band filters $Kp$ and $V$ give similar results. For determination 
of RRab metallicities of LMC, \citet{ata13} used the $V$-band amplitude 
obtained from the $I$-band amplitude given in the 
OGLE-III catalog through theoretical modeling.

Since the $I$-band light curves of the OGLE catalog have large number of data 
points as compared to the $V$-band light curves, the accuracy of the the 
Fourier parameters and hence the determined physical parameters is expected to 
be considerably higher. In the following section, we discuss the 
determination of physical parameters such as metallicities ($[Fe/H]$) and 
absolute magnitudes ($M_{V}$) of RRab stars selected from the OGLE-III 
catalog for the present study.
 
\subsection{[Fe/H] determinations}
\label{iron}
With the recent surge in the ground based  photometric surveys and advent of  
space missions such as CoRoT and Kepler, an unprecedented pool 
of highly accurate and precise RRL light curve data has become available. 
The determination of metallicities of a large number of variable RRL stars 
from the photometric light curves has been possible through the use of various 
empirical relations between the metallicity and light curve parameters 
\citep{jurc96,sand04,smol05}. This is a fast and cheaper 
alternative to the more accurate spectroscopic determinations. Numerous studies in the literature reveal the fact that the parameters determined from 
these empirical relations may not give consistent results for individual stars 
when compared with spectroscopic determinations but, nevertheless, prove to be 
useful when used for large homogeneous samples. For example, in 
order to examine whether there exists a metallicity gradient along the radial 
distance from the centre of a galaxy, the photometrically determined 
metallicities from a large number of RRL stars may provide a clue to the 
answer. They  also allow for a comparative study of 
metallicities among different clusters/galaxies. 

The metallicities of the sample of RRab stars in the present study were 
determined using the empirical relations available in the literature. These 
empirical relations predict the period-metallicity relationship through 
the parameters like amplitude, color, rise time and Fourier phase parameter 
$\phi_{31}$. These empirical period-metallicity relations were made on the 
basis of the fact that these light curve parameters vary with the period 
within the instability strip \citep{sand04}. 

We start with the metallictiy determinations using the relation of  
\citet[hereafter JK96]{jurc96}. The JK96 
$[Fe/H]-P-\phi_{31}$ relation is 
one of the highly used relations for metallicity determinations.  
However, the relation cannot be applied for the estimation of $[Fe/H]$ 
in peculiar stars, such as Blazhko variables and highly evolved stars. In 
order to select a clean sample of RRab stars having `normal-looking' light 
curves, JK96 introduced a compatibility test for 
identifying `peculiar' stars quantified  by the deviations parameter $D_{F}$. 
This is defined as  
\begin{equation}
\label{deviation}
D_{F}=\frac{F_{obs}-F_{calc}}{\sigma_{F_{obs}}},
\end{equation}                                 
where $F_{obs}$ is the observed value of a given parameter, $F_{calc}$ is
the predicted value from other observed parameters and $\sigma_{F}$ is the 
corresponding deviation of various correlations as listed in Table 6 of JK96. 
We have used the compatibility test based on $\phi_{31}$. For 
the $I$-band data, we have used a cut-off limit of $D_{F} = 5$ leaving us with clean sample of 13,095 RRab stars for the determination of $[Fe/H]$ and for 
further analysis. 

The empirical relation of JK96 connecting $\phi_{31}$ in 
the $V$- band, period $P$ and the metallicity $[Fe/H]$ is given by
\begin{equation}
\label{jk96}
[Fe/H]_{JK} = -5.038 - 5.394\, P + 1.345\,\phi_{31},~~\sigma = 0.14.
\end{equation}
To use this relation, we need to convert the $I$-band Fourier parameters to 
those in the $V$-band. In order to make use of Eqn.~\ref{jk96}, we use 
the following relation \citep{ds10}
\begin{equation}
\phi_{31}^{V}=(0.436\pm0.075)+(0.568\pm0.030)\phi_{31}^{I}.
\label{phi31iv}
\end{equation}
The above relation was obtained from a highly accurate set of light curves of 
$29$ RRab variables by \citet{ds10}. The relation exhibits a correlation 
coefficient of $R^2=0.912$, where $R^{2} =1$ corresponds to a perfect 
correlation. It should be noted that the above relation was obtained for the 
cosine Fourier decomposition. A factor of $\pi$ has to be added or 
subtracted in order to convert $\phi_{31}$ from cosine to sine series or 
vice-versa. The metallicity values obtained from Eqn.~\ref{jk96} is in 
\citet{jurc96} scale, which can be transformed into the metallicity scale of 
\citet[hereafter, ZW84]{zw84} using the relation from \citet{jurc95}:
\begin{equation}         
\label{zw_scale}
[Fe/H]_{I}=\frac{[Fe/H]_{JK}-0.88}{1.431}.
\end{equation}
Using the $I$-band data and the corresponding metallicity values, 
\citet{smol05} obtained the following relation involving the  period  and the 
phase parameter $\phi_{31}$ using the sine series
\begin{eqnarray}
\label{smol}
[Fe/H]_{S}=-(3.142\pm0.636)-(4.902P\pm0.375)+~~~~~\nonumber  
\\(0.824\pm0.104)\phi_{31},~~\sigma=0.18. 
\end{eqnarray}                                                        
The above relation given by Eqn.~\ref{smol} is based on the metallicity scale 
of \citet{jurc96}, which is converted into the metallicity scale of ZW84 using
Eqn.~\ref{zw_scale}. Let the metallicity in this new scale be denoted by 
$[Fe/H]_{II}$ as follows:     
\begin{equation}         
\label{sm05}
[Fe/H]_{II}=\frac{[Fe/H]_{S}-0.88}{1.431}.
\end{equation}

\citet{alco2000} showed that the metallicities of RRab stars are linked to the 
period $P$ and $V$-band amplitude of the light through the following empirical 
relation
\begin{equation}
[Fe/H]_{III} = -2.6-8.85[\log(P)-0.15A_{V}],~~\sigma=0.31 
\label{alco2000}
\end{equation}    
where $A_{V}$ is obtained from the following relation of \citet{ds10}
\begin{equation}
A_{V}=0.071(\pm 0.019)+1.500(\pm0.040)A_{I}.
\label{ampiv}
\end{equation}  

\begin{figure}
\includegraphics[width=0.5\textwidth,keepaspectratio]{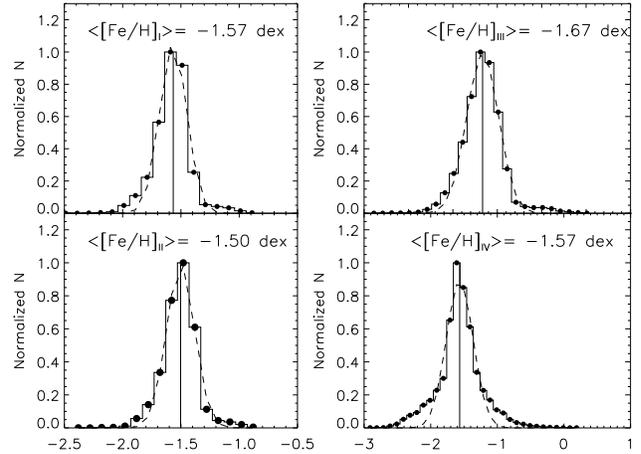}
\caption{Metallicity distribution of $13,095$ RRab stars selected for the 
present analysis using the four empirical relations as described in the text.}
\label{feh_comp}
\end{figure}

Another estimate of $[Fe/H]$ can be obtained from 
$[Fe/H]-\log{P}-\phi_{31}$ relation of \citet{sand04} given by 
\begin{equation}
[Fe/H]_{IV} =-6.025-7.012\log{P}+1.411\phi_{31},~~\sigma=0.16.
\label{s04}
\end{equation} 
The above relation is based on the cosine series. The intercept term in this relation has an uncertainty of $0.023$, $\phi_{31}$ and $\log{P}$ have respective 
uncertainties of $0.014$ and $0.071$. It should be noted that the determined 
metallicity values using the above relations do not yield the same result on a 
star-to-star basis. We have fitted a three parameter Gaussian function to each 
metallicity distribution of the present sample of $13,095$ RRab stars. The 
following peak values are obtained: $<[Fe/H]_{I}> = -1.57 \pm 0.12$ dex, 
$<[Fe/H]_{II}> = -1.50\pm0.12$ dex, $<[Fe/H]_{III}> = -1.67\pm0.18$ dex and 
$<[Fe/H]_{IV}>= -1.57\pm0.20$ dex. In the determination of these peak values, a 
constant bin size of $0.2$ dex has been used.
Fig.~\ref{feh_comp} shows the metallicity distribution of $13,095$ RRab stars
obtained using the above four empirical relations.    
\begin{figure}
\includegraphics[width=0.5\textwidth,keepaspectratio]{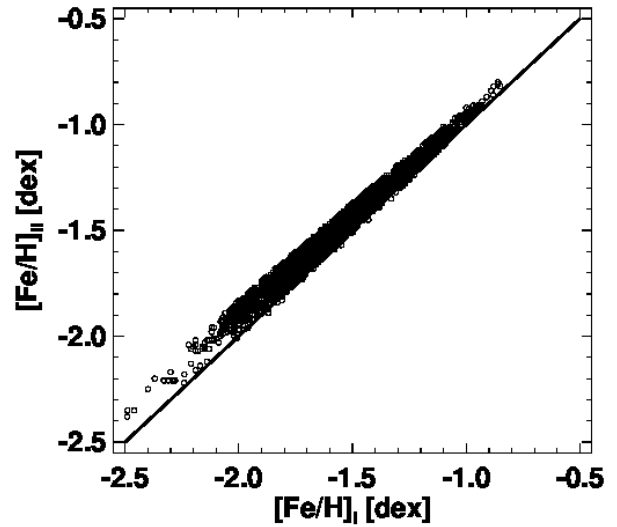}
\caption{Comparison of metallicities of $13,095$ stars determined using the 
\citet{smol05} relation for the $I$-band ($[Fe/H]_{II}$) and those determined 
from the $JK96$ relation using the calibration relation of Eqn.~\ref{phi31iv} 
in the present study ($[Fe/H]_{I}$).}
\label{fesmolec}.  
\end{figure}
A scatter plot of the metallicity values of $13,095$ RRab stars 
obtained using the \citet{smol05} is shown in Fig.~\ref{fesmolec} against 
the metallicity values using the JK96 relations after $\phi_{31}^{I}$ has been 
converted into the corresponding $\phi_{31}^{V}$ using the calibration 
relation given by Eqn.~\ref{phi31iv}. The metallicity values obtained using 
the \citet{smol05} relation are on an average higher by $0.07$ dex. We adopt 
the metallicity values $[Fe/H]_{I}$  in the analysis that follows and 
denote these as $[Fe/H]$.

\begin{figure}
\includegraphics[width=0.5\textwidth,keepaspectratio]{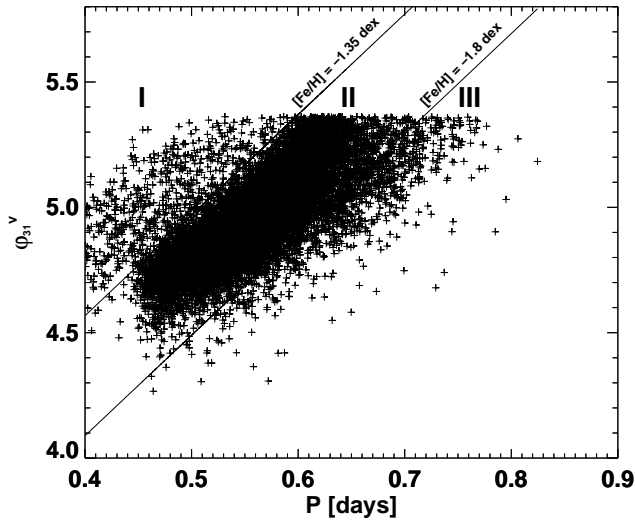}
\caption{Representation of $13,095$ stars on the $P-\phi_{31}$ plane. The 
diagram is clearly separated into three regions marked by $I, II$ and $III$ 
which correspond to the stars with $[Fe/H] \ge -1.35$ dex, 
$-1.80 < [Fe/H]< -1.35$ dex and $[Fe/H] \le -1.80$ dex, respectively.}
\label{met1}. 
\end{figure}

\begin{figure}
\includegraphics[width=0.5\textwidth,keepaspectratio]{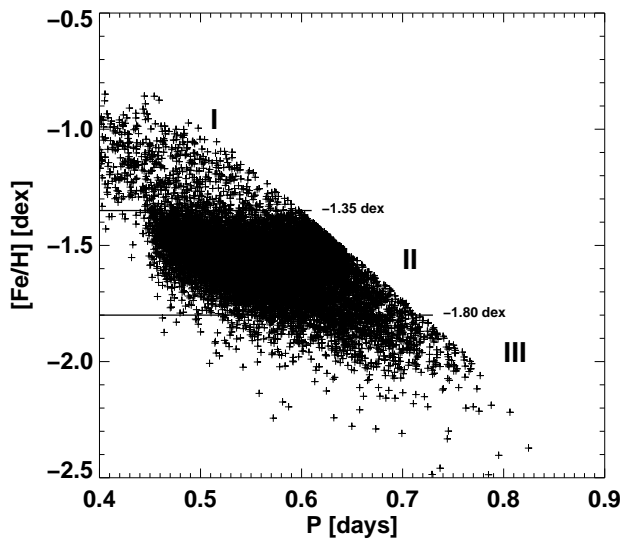}
\caption{Representation of $13,095$ stars on the $P-[Fe/H]$ plane. The 
three regions having three different metallicity groups as in Fig.~\ref{met1} 
are marked by $I, II$ and $III$, respectively.}
\label{met2}.
\end{figure}

Fig.~\ref{met1} shows the parameter space of $\phi_{31}^{V}$ versus $P$. An 
apparent random scatter of the data points is quite discernible. The scatter is
in fact strongly correlated with the metallicity. The strong separation of three
metallicity groups is quite evident from the figure. The segregation of data 
points on the $\phi_{31}^{V}-P$ diagram allows them to be divided into 
three metallicity groups, namely, $I, II$ and $III$ with metallicity values in 
the range $[Fe/H] \ge -1.35$ dex (metal-rich), $-1.80 < [Fe/H] < -1.35$ 
dex (metal-poor), and $[Fe/H] \le -1.80$ dex (extremely metal-poor), 
respectively as obtained from the data. Fig.~\ref{met2} shows the three groups
in the $P-[Fe/H]$ plane. The number of stars in each of the 
groups $I, II$ and $III$ are  $659$, $11406$ and $1030$, respectively. The 
mean metallicities of these three groups are found to be 
$-1.20 \pm 0.12$ dex ($I$), $-1.57 \pm 0.10$ dex ($II$) and $-1.89 \pm 0.09$ 
dex ($III$). 
\subsection{Absolute magnitudes and distances}
The empirical relation to estimate the absolute magnitudes ($M_{V}$) of RRab 
stars in terms of the period ($P$) and the Fourier coefficients ($A_{1}$ and 
$A_{3}$) is given by \citet{kova01}:
\begin{equation}
\label{abs_mag}
M_{V}=-1.876\log{P}-1.158A_{1}+0.821A_{3}+K,
\end{equation}
Using the distance modulus of $18.50$ mag for the LMC \citep{free01}, 
\citet{ferr10} found $K=0.41$. We use $K=0.41$ in Eqn.~\ref{abs_mag} to 
estimate the absoulte magnitude of $13,095$ RRab stars of the LMC. Mean 
values of absolute magnitudes of the three metallicity groups are: 
$0.70 \pm 0.08$ mag ($I$), $0.59 \pm 0.06$ mag ($II$) and $0.49 \pm 0.08$ mag ($III$).  
The absolute magnitudes are converted to luminosities using the standard 
relation
\begin{equation}
\log{\left(\frac{L}{L_{\sun}}\right)}=\left(4.75-(M_{V}+BC_{V})\right).
\end{equation}  
Here, we use $M_{bol}(\text{Sun}) = +4.75$. The bolometric correction 
($BC_{V}$) is calculated using the equation \citep{sand90} 
\begin{equation}
BC_{V}=0.06+0.06[Fe/H].
\end{equation}  
Once the absolute magnitudes of the RRL stars are obtained, the 
intensity-weighted mean magnitude ($<V>$) were used to derive their 
distance moduli. The values of $<V>$ have been calculated using Eqn.~15 
of \citet{ds10}. The values of $<I>$ were calculated following 
\citet{saha90}. Mean values of $<I>$ for each of these groups are 
$19.07 \pm 0.27$ mag, $18.83 \pm 0.24$ mag and $18.22 \pm 0.25$ mag, whereas 
the corresponding mean values of $<V>$ are $19.63 \pm 0.27$ mag, 
$19.39 \pm 0.24$ mag and $19.22 \pm 0.25$ mag, respectively.

\begin{figure}
\includegraphics[width=0.5\textwidth,keepaspectratio]{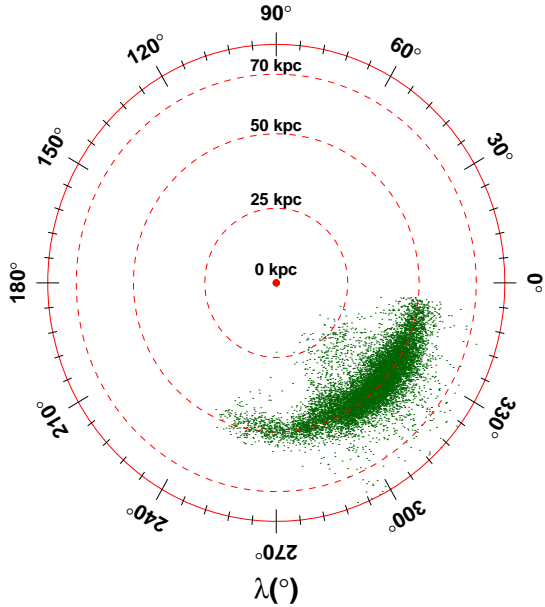}
\caption{Polar plot of the distance distribution of LMC RRab stars in 
ecliptic coordinates. The dashed lines are set at distances $D = 25, 50, 70~\text{kpc}$, respectively.}
\label{polar}.  
\end{figure}

In order to determine the interstellar extinction 
$\mathbb{A}_{V}$, we use the following relation \citep{schl98} 
\begin{equation}
\mathbb{A}_{V}=3.24(E(V-I)/1.4),
\end{equation}
where the reddening values $E(V-I)$ are calculated from the LMC extinction map 
based on the OGLE-III RRL stars by \citet{pejcha09}. Mean distances to 
each of the three metallicity groups are: $52.25\pm6.20$ kpc ($I$), $49.36\pm 5.30$ kpc ($II$) and $47.87 \pm 5.25$ kpc ($III$). Table~\ref{vpar} shows the 
various parameters determined for the $13,095$ RRab stars in this study. $\mu$ 
and $D$ represent the distance modulus in mag and distance in kpc, 
respectively. Fig.~\ref{polar} shows polar plot of the distance distribution 
of LMC RRab stars in ecliptic coordinates. The dashed lines are set at distances
D = 25, 50, 70 kpc, respectively.            

\begin{table*}
\caption{ Various parameters extracted from Fourier coefficients for $13,095$ 
RRab variables. Errors represent the uncertainties in the Fourier parameters. 
The complete table is available in the electronic version of this paper.}
\begin{center}
\label{vpar}
\scalebox{0.9}{
\begin{tabular}{cccccccccccccccc}
\\
\hline
OGLE ID & $[Fe/H]~[\text{dex}]$ &$M_{V}~[\text{mag}]$& $\log(L/L_{\odot})$&$<I>~[\text{mag}]$&$<V>~[\text{mag}]$&$\mu~[\text{mag}]$&
$D~[\text{kpc}]$ \\
~~~~~(1)&(2)&(3)&(4)&(5)&(6)&(7)&(8)\\
\hline
OGLE-LMC-RRLYR-23488&-1.09$\pm$ 0.11& 0.80$\pm$0.008& 1.58$\pm$ 0.00&     19.12&     19.68&18.61$\pm$ 0.04&  52.61$\pm$   0.40\\
OGLE-LMC-RRLYR-20444&-1.17$\pm$ 0.12& 0.81$\pm$0.009& 1.58$\pm$ 0.00&     19.33&     19.89&18.66$\pm$ 0.07&  53.98$\pm$   0.74\\
OGLE-LMC-RRLYR-17769&-1.12$\pm$ 0.09& 0.76$\pm$0.003& 1.60$\pm$ 0.00&     19.06&     19.61&18.53$\pm$ 0.09&  50.93$\pm$   0.92\\
OGLE-LMC-RRLYR-21760&-0.96$\pm$ 0.13& 0.80$\pm$0.010& 1.58$\pm$ 0.01&     19.48&     20.03&18.72$\pm$ 0.06&  55.52$\pm$   0.63\\
OGLE-LMC-RRLYR-11295&-1.15$\pm$ 0.13& 0.74$\pm$0.007& 1.61$\pm$ 0.00&     19.10&     19.66&18.69$\pm$ 0.04&  54.83$\pm$   0.41\\
OGLE-LMC-RRLYR-03670&-0.98$\pm$ 0.14& 0.84$\pm$0.006& 1.56$\pm$ 0.00&     19.34&     19.89&18.76$\pm$ 0.07&  56.40$\pm$   0.84\\
OGLE-LMC-RRLYR-13862&-1.13$\pm$ 0.09& 0.78$\pm$0.004& 1.59$\pm$ 0.00&     19.12&     19.68&18.69$\pm$ 0.07&  54.63$\pm$   0.77\\
OGLE-LMC-RRLYR-23324&-0.95$\pm$ 0.12& 0.79$\pm$0.009& 1.58$\pm$ 0.00&     19.23&     19.78&18.68$\pm$ 0.04&  54.51$\pm$   0.45\\
OGLE-LMC-RRLYR-01648&-1.10$\pm$ 0.11& 0.79$\pm$0.008& 1.58$\pm$ 0.00&     19.19&     19.74&18.68$\pm$ 0.04&  54.49$\pm$   0.40\\
OGLE-LMC-RRLYR-07370&-1.33$\pm$ 0.13& 0.78$\pm$0.004& 1.60$\pm$ 0.00&     19.13&     19.68&18.45$\pm$ 0.18&  48.99$\pm$   1.75\\
OGLE-LMC-RRLYR-20673&-1.21$\pm$ 0.13& 0.75$\pm$0.009& 1.61$\pm$ 0.00&     19.17&     19.73&18.68$\pm$ 0.06&  54.36$\pm$   0.64\\
\hline
\hline
\end{tabular}
}
\end{center}
\end{table*}  

\section{Structure of the LMC}
\label{structure}
\begin{figure}
\includegraphics[width=0.5\textwidth,keepaspectratio]{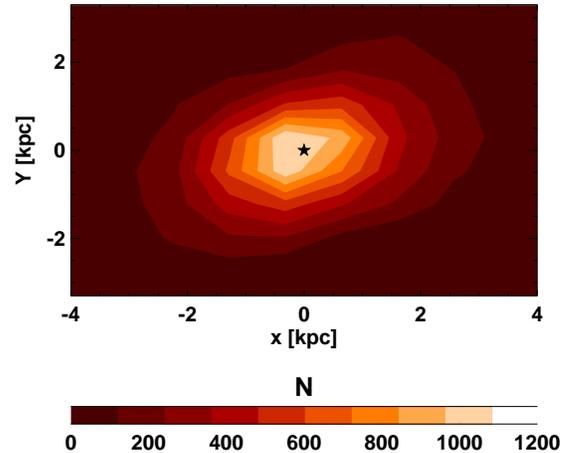}
\caption{Two dimensional number density of $13,095$ RRab stars is shown as a
contour map. The star symbol denotes the location of the centroid for the 
present sample.} 
\label{contour}
\end{figure}

The Cartesian coordinates corresponding to each star can be obtained using 
the $RA (\alpha), Dec (\delta)$ and the distance $D$ in kpc.   
Let us consider the Cartesian coordinate system $(x,y,z)$ which 
has the origin at the center of the LMC at $(\alpha,\delta,D)=(\alpha_{0},\delta_{0},D_{0})$. The $z$-axis is pointed towards the observer, the $x$-axis 
is antiparallel to the $\alpha$-axis and the $y$-axis is parallel to the 
$\delta$-axis.	$D_{0}$ is the distance between the center of the LMC and the 
observer while $D$ is the observer-source distance and $(\alpha_{0},\delta_{0})$ are 
the equitorial coordinates of the center of the LMC. The $(x,y,z)$
coordinates are obtained using the transformation equations  
\citep{vand101,wein01}:
\begin{eqnarray}
\label{proj}
x=-D\sin(\alpha-\alpha_{0})\cos{\delta}, \nonumber \\
y=D\sin{\delta}\cos{\delta_{0}}-D\sin{\delta_{0}}\cos{(\alpha-\alpha_{0})}\cos{\delta}, \\
z=D_{0}-D\sin{\delta}\sin{\delta_{0}}-D\cos{\delta_{0}}\cos{\alpha-\alpha_{0}}\cos{\delta} \nonumber.
\end{eqnarray}
The coordinate system of the LMC disk $(x^{\prime},y^{\prime},z^{\prime})$ is 
the same as the orthogonal system $(x,y,z)$, except that it is rotated around 
the $z$-axis by position angle $\theta$ counterclockwise and around the 
new $x$-axis by the inclination angle $i$ clockwise. 
The coordinate transformations can be written as  \citep{vand101,wein01}:
\begin{eqnarray*}
x^{\prime}=x\cos{\theta}+y\sin{\theta}, \\
y^{\prime}=-x\sin{\theta}\cos{i}+y\cos{\theta}\cos{i}-z\sin{i}, \\
z^{\prime}=-x\sin{\theta}\sin{i}+y\cos{\theta}\sin{i}+z\cos{i}.
\end{eqnarray*}
The two dimensional density contours of the distribution of the LMC RRab stars 
are shown in Fig.~\ref{contour}. The star symbol denotes the location of the 
centroid of the present sample.      
\subsection{Vertical distribution of the stars}
\begin{figure}
\includegraphics[width=0.5\textwidth,keepaspectratio]{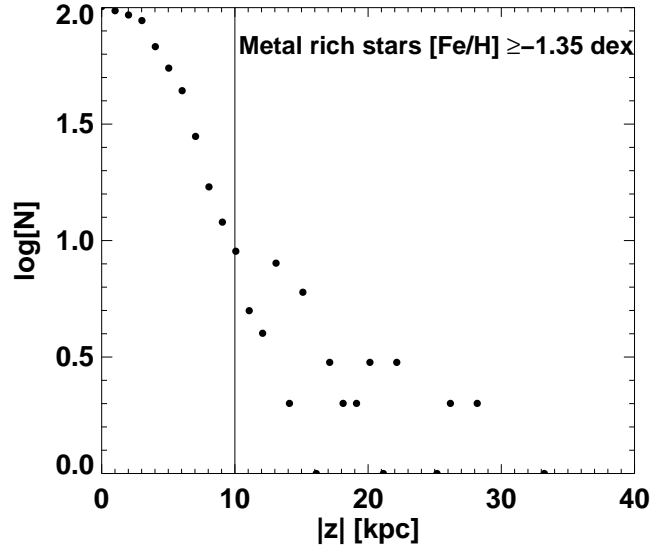}
\caption{ Distribution of metal rich ($[Fe/H] \ge -1.35$ dex) RRab stars of the 
LMC as a function of vertical $|z|$ distance from the galactic plane.}
\label{zdist_feh1}
\end{figure}

\begin{figure}
\includegraphics[width=0.5\textwidth,keepaspectratio]{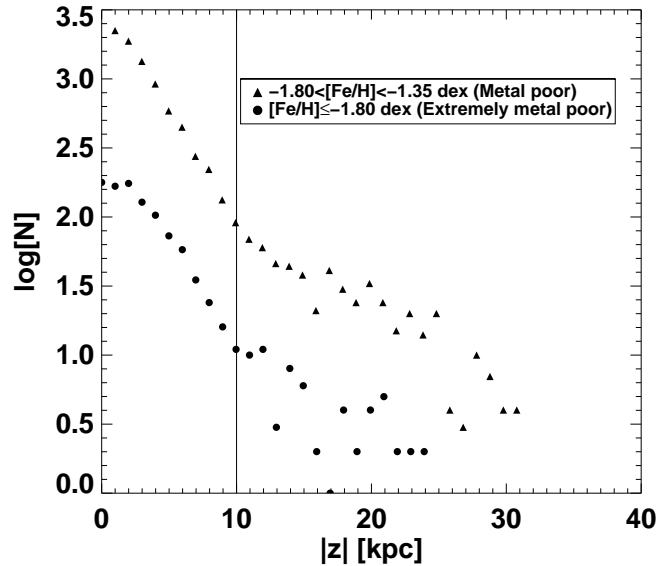}
\caption{Distribution of metal poor ($-1.80 < [Fe/H] < -1.35$ dex) and 
extremely metal rich ($[Fe/H] \le -1.80$ dex) RRab stars of the LMC as a 
function of vertical $|z|$ distance from the galactic plane.}
\label{zdist_feh2}
\end{figure}

In this section, we study the vertical $|z|$ distribution of the LMC RRab 
stars. In order to see how the metal poor  and metal rich RRab stars are 
distributed in the vertical, we divide the stars into three metallcity groups, 
viz, $I, II$ and $III$ as described in section~\ref{iron}. We further use a 
bin size of $1$ kpc and count the number of stars falling in each bin. 
The resulting distribution of metallicity group $I$ (metal rich stars) is 
shown in Fig.~\ref{zdist_feh1}. From the figure, we can see that the stars 
of this group are mainly distributed  upto a distance of 
$\sim 10$ kpc, after which the distribution becomes uniform. The number of 
metal rich stars lying below $z = 10$ kpc are $639~(\sim 97.0\%)$ while lying 
above it are $20~(\sim 3.0\%)$ only, out of a total of $659$. On the other hand, 
the distribution of the RRab stars belonging to metallicity groups $II$ 
(metal poor) and $III$ (extremely metal poor) are shown in 
Fig.~\ref{zdist_feh2}. Upto a 
distance of $10$ kpc, the number of stars belonging to the groups $II$ and 
$III$ are $11,015$ and $973$, out of a total of $11,406$ and $1030$, respectively. 
These account for $96.57\%$ and $94.47\%$, respectively of the total.  The number 
of stars in each of these groups decrease with increasing distance from the 
galactic plane. Also, it can be seen that number of metal poor stars are very 
large as compared to the metal rich stars within a distance of $10$ kpc. It 
is also found that more than $94\%$ ($12,340$) of the total number of RRab stars 
lie within $|z| =10$ kpc. Less than $10\%$ of the RRab stars are located 
beyond a $|z|$ distance of $10$ kpc with majority of them being metal poor 
($[Fe/H] <  -1.35$ dex). This implies that the RRLs in the LMC belong to two 
different structures, one with smaller scale height tracing the disk and the 
other with larger scale height tracing the inner halo of the LMC. 

\begin{figure}
\includegraphics[width=0.5\textwidth,keepaspectratio]{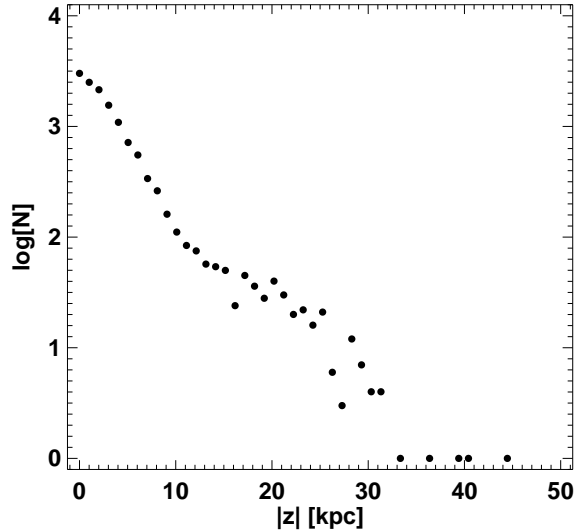}
\caption{Distribution of all RRab stars of the LMC as a function of 
vertical $|z|$ distance from the galactic plane.}
\label{zdist}
\end{figure}

\begin{figure}
\includegraphics[width=0.5\textwidth,keepaspectratio]{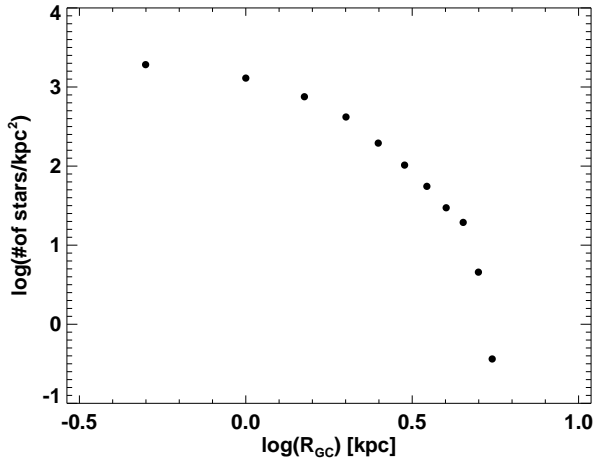}
\caption{Radial number density distribution of $13,095$ RRab stars as a 
function of galactocentric distance in logarithmic scale.}
\label{radial_dens}
\end{figure}

Since a  majority of the metal poor RRab stars lie within $|z| = 10$ kpc 
which traces the disk, we may conclude that the disk of the LMC has been 
formed much earlier than the extended halo of the LMC. This confirms the 
findings of \citet{as09} that RRLs in the inner LMC belong to two different 
populations tracing the disk and the inner halo. The spatial distribution of the 
LMC RRab stars and their  metallicities indicate 
that the majority of the old and metal-poor LMC field stars lie in a disk and 
not in a spheroid. Fig.~\ref{zdist} shows the $|z|$ distribution of all RRab 
stars of the LMC selected in the present study. The radial number density 
profile of all the RRab stars as a function of galactocentric distance is shown in 
Fig.~\ref{radial_dens}. The radial number density profile is obtained by 
projecting the radial number density of RRabs in concentric rings around the 
centroid of the LMC.          
\subsection{Inclination angle ($i$) and position angle of the line of nodes 
($\theta_{lon}$)}
Once we have the distribution of RRab distances as determined above, we apply 
the least square plane fitting method and the method of principal axes 
transformation using the moment of inertia tensor in order to get an estimate 
of the viewing angles of the LMC. In order to obtain the viewing angles of the 
LMC disk, viz., the inclination $i$ and position angle of the line of nodes 
$\theta_{lon}$, the distances are projected on coordinate axes $(x,y,z)$. Once 
the $x$, $y$ and $z$ coordinates are obtained using Eqn.~\ref{proj}, we 
apply a plane fit equation of the form \citep{niko04} 
\begin{equation}
z_{i}=ax_{i}+by_{i}+c,~~i=1,2,\dots,N.
\end{equation}                    
The fit to $z_{i}$ is obtained as a function of $(x_{i}, y_{i})$. From 
the coefficients of the plane fit solution, the inclination ($i$) and position 
angle of the line of nodes ($\theta$) can be calculated using the following 
formula \citep{niko04} 
\begin{eqnarray}
i=\arccos{(\frac{1}{\sqrt{(1+a^{2}+b^{2})}})},\\
\theta=\arctan{(-\frac{a}{b})}+{\rm sign}{(b)}\frac{\pi}{2}.
\end{eqnarray} 
The plane fitting procedure has been carried out using $|z|=10$ kpc ($12,340$ RRab stars), taken as  the boundary of the LMC disk obtained from the vertical 
$z$ distribution of the metal rich RRab stars. The values 
obtained for the LMC disk are $i=36^{\circ}.43$ and $\theta_{lon}=149^{\circ}.08$. 
    
Now, we model the observed population of the RRab stars in LMC by a triaxial 
ellipsoid upto a $|z|$ distance of $10$ kpc. The properties of the ellipsoid 
can be obtained from the moment of inertia tensor using the principal axes 
transformation. In order to apply the principal axes transformation, we 
first construct the covariance matrix (inertia tensor) from the $[x(i), y(i), 
z(i), i=1,\dots,N]$ distribution of the RRab stars. To construct the 
covariance matrix, we first determine the centroid of the distribution using 
\citet{bestfit} 
\begin{eqnarray}
\bar{x}=\sum_{i=1}^{N}x(i)/N, \nonumber \\
\bar{y}=\sum_{i=1}^{N}y(i)/N, \\
\bar{z}=\sum_{i=1}^{N}z(i)/N, \nonumber
\end{eqnarray}                  
and then construct the  covariance matrix called the moment of inertia tensor as 
\citep{bestfit,huan11}
\begin{equation}
I= \begin{bmatrix}
~~I_{xx}&-I_{xy}&-I_{xz} \\
-I_{yx}&~~I_{yy}&-I_{yz}\\
-I_{zx}&-I_{zy}&~~I_{zz}
\end{bmatrix},
\end{equation} 
where
\begin{eqnarray}
I_{xx}=\Theta_{yy}+\Theta_{zz}, \nonumber \\
I_{yy}=\Theta_{xx}+\Theta_{zz}, \nonumber \\
I_{zz}=\Theta_{xx}+\Theta_{yy}, \nonumber \\
I_{xy}=I_{yx}=\Theta_{xy}, \\ 
I_{xz}=I_{zx}=\Theta_{xz}, \nonumber \\
I_{yz}=I_{zy}=\Theta_{yz}, \nonumber
\end{eqnarray}
and 
\begin{eqnarray}
\Theta_{xx}=\frac{1}{N}\sum_{i=1}^{N}\left(x(i)-\bar{x}\right)^{2}, \nonumber \\
\Theta_{yy}=\frac{1}{N}\sum_{i=1}^{N}\left(y(i)-\bar{y}\right)^{2},\nonumber \\
\Theta_{zz}=\frac{1}{N}\sum_{i=1}^{N}\left(z(i)-\bar{z}\right)^{2},\nonumber \\
\Theta_{xy}=\frac{1}{N}\sum_{i=1}^{N}\left(x(i)-\bar{x}\right)\left(y(i)-\bar{y}\right), \\
\Theta_{xz}=\frac{1}{N}\sum_{i=1}^{N}\left(x(i)-\bar{x}\right)\left(z(i)-\bar{z}\right), \nonumber \\
\Theta_{yz}=\frac{1}{N}\sum_{i=1}^{N}\left(y(i)-\bar{y}\right)\left(z(i)-\bar{z}\right). \nonumber 
\end{eqnarray}
When the axes of the coordinate frame ($x,y,z$) are selected such that 
$I_{xy}=I_{xz}=I_{yz}=0$, we have the principal axes of inertia of 
the system. The corresponding moments of inertia $I_{xx}, I_{yy}$, and $I_{zz}$
are the principal moments of inertia \citep{huan11}. It can be seen 
that the covariance matrix $I$ is real and symmetric. Let  the matrix $I$ have 
the eigenvalues ($\lambda_{1}>\lambda_{2}>\lambda_{3}$) and 
$\{\vec{e_{1}^{\prime}},\vec{e_{2}^{\prime}},\vec{e_{3}^{\prime}}\}$ be the 
corresponding eigenvectors normalized to unity. In order to obtain the 
principal axes of the coordinate frame, we need to diagonalize the symmetric 
covariance matrix. The eigenvectors of the matrix $I$ can be used to form 
another matrix $T$ such that the matrix $T^{-1}IT$ is a diagonal matrix. The 
diagonal elements of the diagonal matrix are the eigenvalues of $I$ 
\citep{pres02,tang06,bestfit}. Here $T$ is the matrix formed with the 
eigenvectors of $I$ as the column vectors. It can be shown that the matrix $T$ 
is an orthogonal matrix and diagonalizes the covariance matrix $T$ 
\citep{pres02,tang06,huan11}. The new coordinate axes in which the matrix $I$ 
is diagonal are known as the principal axes \citep{pres02,tang06}. The 
diagonalization 
\begin{equation}
I^{\prime}=T^{-1}IT=T^{\dag}IT,
\end{equation}
yields
\begin{equation}
I^{\prime}=
\begin{bmatrix}
\lambda_{1}&0&0 \\
0&\lambda_{2}&0 \\
0&0&\lambda_{3}
\end{bmatrix}.
\end{equation}                
The three eigenvalues ($\lambda_{1}>\lambda_{2}>\lambda_{3}$) correspond 
to non-zero diagonal terms of the inertia tensor in the new coordinate system  
and are called the principal moments of inertia of the system. The 
eigenvectors corresponding to the three eigenvalues represent three orthogonal 
axes in the new coordinate system and are called the principal axes of the new 
coordinate system \citep{pres02,tang06}. 
The transformation matrix or the rotation matrix which carries out the transformation $T:R^{3}\rightarrow {R^{\prime}}^{3}$ 
is given by 
\[T(\vec{e})=\vec{e^{\prime}},\]
where $\vec{e}=(e_{1},e_{2},e_{3})^{T}$ and  $\vec{e^{\prime}}=(e_{1}^{\prime},e_{2}^{\prime},e_{3}^{\prime})^{T}$, where ${\vec{e_{i}}, 
\vec{e_{i}^{\prime}}}$ are the basis vectors in the Cartesian coordinate 
systems $(x,y,z)$ and the new coordinate system, respectively. The basis vectors 
${\vec{e_{i}}}$ and ${\vec{e_{i}^{\prime}}}$ are related by an orthogonal 
tensor {\bf T} through the equations given below
\begin{equation} 
\vec{e_{i}^{\prime}}=\mathbf{T}\vec{e_{i}}=\sum_{j=1}^{3}T_{ij}\vec{e_{j}}
\end{equation}
That is, 
\begin{eqnarray}
\vec{e_{1}^{\prime}}=T_{11}\vec{e_{1}}+T_{12}\vec{e_{2}}+T_{13}\vec{e_{3}}, \\
\vec{e_{2}^{\prime}}=T_{21}\vec{e_{1}}+T_{22}\vec{e_{2}}+T_{23}\vec{e_{3}}, \\
\vec{e_{3}^{\prime}}=T_{31}\vec{e_{1}}+T_{32}\vec{e_{2}}+T_{33}\vec{e_{3}}, \\
\end{eqnarray}
where 
\begin{align}
T_{ij}T_{kj}=T_{ji}T_{jk}=\delta_{ik}
\Rightarrow TT^{\dag}=TT^{\dag}=1
\end{align}
It may be noted that
\begin{align}
T_{ij}=\cos{(\vec{e_{i}},\vec{e_{j}^{\prime}})}.
\end{align} 
The matrix formed from these direction cosines, i.e., the matrix
\begin{align}
\begin{bmatrix} 
T_{11} & T_{21}&T_{31} \\ 
T_{12} & T_{22}&T_{32} \\
T_{13} & T_{23}&T_{33} \\
\end{bmatrix},
\end{align}  
is called the transformation matrix between 
$\{\vec{e_{1}},\vec{e_{2}},\vec{e_{3}}\}$ and 
$\{\vec{e_{1}^{\prime}},\vec{e_{2}^{\prime}},\vec{e_{3}^{\prime}}\}$.
The transformation matrix or the rotation matrix consists of the eigenvectors 
of $I$ as the column vectors. The transformation matrix $T$ describes the 
spatial directions or the orientation of the ellipsoid with respect to the 
local coordinate  system $(x,y,z)$ \citep{pres02,tang06,bestfit}.    

\begin{figure}
\includegraphics[width=0.5\textwidth,keepaspectratio]{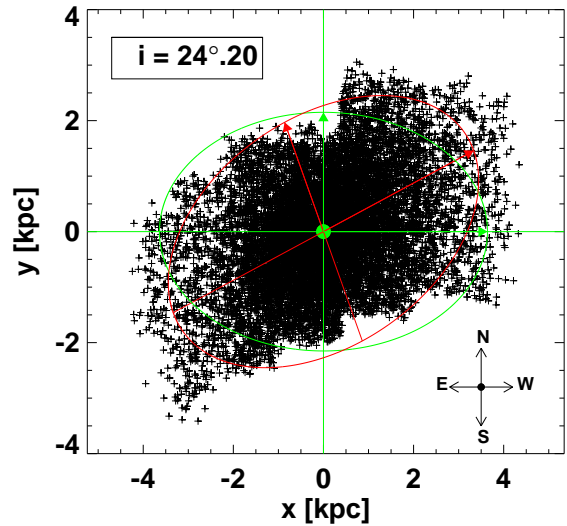}
\caption{The value of the inclination angle as obtained from the principal 
axes transformation of the moment of inertia tensor is shown. The filled green 
circle denotes the origin ($0,0$) of the distribution. $\rm N,E,S,W$ represent 
the  north, east, south and west directions, respectively.}
\label{incl}
\end{figure}

\begin{figure}
\includegraphics[width=0.5\textwidth,keepaspectratio]{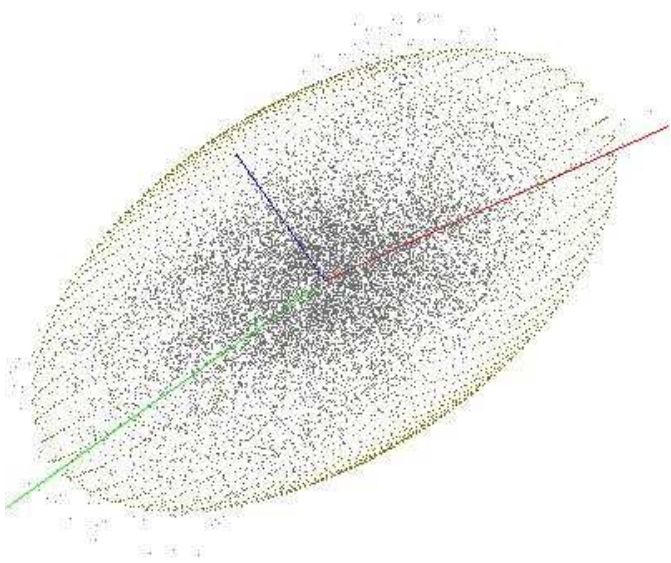}
\caption{Best fit-ellipsoid obtained from the principal axis transformation of 
the moment of inertia tensor constructed from the $x,y,z$ distributions of the 
RRab stars in the LMC.}
\label{ellipsoid}
\end{figure}

Applying the above method  to the distribution of the RRab stars in the 
coordinate system, we obtain the eigenvalues $\lambda_{1}= 14.85, 
\lambda_{2}=13.46$ and $\lambda_{3}=2.92$. From the transformation matrix, we 
estimate the inclination angle and position angle of the line of nodes as 
$24^{\circ}.20$ and $176^{\circ}.01$ respectively. The lengths of the 
semi-axes ($S_{i}$) of the best-fit ellipsoid are obtained using \citep{bestfit}
\begin{equation}
S_{i}=\sqrt{\frac{5}{2}(\lambda_{j}+\lambda_{k}-\lambda_{i})},~~\text{for}~~i \ne j \ne k. 
\end{equation}
where $S_{1}> S_{2}> S_{3}$ are the major-axis and two minor axes, 
respectively. The lengths of the axes are estimated as: 
$S_{1} = 7.97~\text{kpc}, S_{2} = 3.28~\text{kpc}$ and  
$S_{3}= 1.96~\text{kpc}$. The eccentricity of the LMC disk is found to be 
$e = 0.41$. The value of the inclination angle as determined from the best-fit ellipsoid in the $(x,y)$- plane is shown in Fig.~\ref{incl}.      
The best-fit ellipsoid to the three-dimensional $(x,y,z)$ distribution is shown
in Fig.~\ref{ellipsoid}.

\subsection{Dependence of viewing angles on the choice of the LMC center}
In this section, we study the effect of the choice of the LMC center on the 
determination of the viewing angles of the LMC disk. The LMC does not have a
unique, well-defined viewing angle \citep{vand101}. Thus, the choice of the 
LMC center taken as the centroid in this study is to some extent arbitrary. 
However, we find that this has no influence on the accuracy of the results 
obtained. It does not lead to statistically different results for the best 
fitting viewing angles $(i, \theta_{lon})$. In order to corroborate this fact, 
we consider the following five choices of the coordinate origin: 
$(1)$ optical center, $(\alpha_{0},\delta_{0}) = (79^{\circ}.91, -69^{\circ}.45, \text{J2000})$ from \citet{deva72}, $(2)$ center obtained from the 
distribution of novae, $(\alpha_{0},\delta_{0}) = (80^{\circ}.07, -69^{\circ}.06, \text{J2000})$ from \citet{vand88}
$(3)$ center defined by $H I$ rotation map, 
$(\alpha_{0},\delta_{0}) = (79^{\circ}.4, -69^{\circ}.03, \text{J2000})$ from
\citet{kim98} 
$(4)$ center of carbon stars outer isopleths, $(\alpha_{0},\delta_{0}) = (82^{\circ}.25, -69^{\circ}.5, \text{J2000})$ from \citet{vand101} and (5) geometric center obtained from Cepheids, 
$(\alpha_{0},\delta_{0}) = (80^{\circ}.40, -69^{\circ}.00, \text{J2000})$ from 
\citet{niko04}. The parameters estimated using the above centers are listed in 
Table~\ref{center}. We find that there are meagre changes in the best-fitted 
viewing angles $(i,\theta_{lon})$ of the LMC disk on the choice of the origin.    
\begin{table*}
\begin{center}
\caption{Estimates of the viewing angles of the LMC obtained in this  study 
for various choices of the LMC center in the reference with different types of 
tracers used given below. The viewing angle parameters in this study have been 
estimated using the principal axes transformation method}
\label{center}
\begin{tabular}{|l|c|c|c|} \\ \hline\hline
Reference& $i$ & $\theta_{lon}$ & Tracers used \\
\hline \hline
\citet{deva72}&$24^{\circ}.62$&$177^{\circ}.16$&Yellow light isophotes \\\hline
\citet{vand88}&$23^{\circ}.50$&$175^{\circ}.70$&Novae \\\hline
\citet{kim98}&$25^{\circ}.24$&$176^{\circ}.38$& HI \\\hline
\citet{vand101}&$22^{\circ}.43$&$175^{\circ}.27$&AGB stars \\\hline
\citet{niko04}&$22^{\circ}.25$&$175^{\circ}.22$& Cepheids\\\hline
Centroid of the present sample &$24^{\circ}.20$&$176^{\circ}.01$& RRab stars\\\hline\hline
\end{tabular}
\end{center}
\end{table*}       
\subsection{Comparison with previous studies}
There have been a number of previous studies on the geometrical parameters of 
the LMC using different tracers. Use of different types of tracers in order to 
disseminate the geometry yields wide range of values of the LMC disk 
inclination and the position angle of the line of nodes. As listed in the 
Table~3.5 of \citet{westerlund97}, the inclination values of the LMC disk 
range from $27$ to $48^{\circ}$ and the position angles of the line of nodes 
vary from $168$ to $208^{\circ}$ depending on the tracers used. The various 
other determinations of the LMC geometry obtained in the last $10$ years are 
listed in Table~\ref{geom}.

A substantial number of studies which have attempted to find the viewing 
angles of the LMC disk using different types of tracers are tabulated in 
Table~1 of \citet{as13}. The inclination values of the LMC disk in most of 
these studies are consistent with each other within the error bars. But, there 
is a lot of discrepancy when it comes to the determination of the position 
angles of the line of nodes. In this study of the LMC using the RRab stars 
from the OGLE-III database, we found an inclination $(i)$ =$24^{\circ}.20$ and 
position angle of the line of nodes $(\theta_{lon}) = 176^{\circ}.01$ from the 
method of principal axes transformation using the moment of inertia tensor. 
Using different types of tracers in the LMC, \citet{wein01} found the 
inclination of the LMC to the line of sight $i=22$ to $29^{\circ}$  and 
position angle of line of nodes $\theta_{lon} = 168$ to $173^{\circ}$ obtained 
by fitting each population by two different models: a thin exponential disk 
and a spherical power-law model. The values of the viewing angles 
$(i,\theta_{lon}) = (24^{\circ}.20, 176^{\circ}.01)$ 
derived here from the principal axes transformation method are consistent with 
the values of \citet{wein01} and the values listed in \citet{westerlund97}.       

On the other hand, using the simple plane fitting procedure on the same 
dataset here yields $i = 36^{\circ}.43$ and $\theta_{lon} = 149^{\circ}.08$. 
The variation in the values of these geometrical parameters of the the LMC can 
be attributed to the highly complicated structure of the LMC as well as 
different methodology adopted in their determinations.         
\begin{table*}
\caption{Various values of the viewing angle parameters $(i,\theta_{lon})$ of 
the LMC disk obtained in the last $10$ years}
\label{geom}
\begin{tabular}{lccc}
\hline \hline
Reference&$i$&$\theta_{lon}$&Tracers used \\ \hline
\citet{niko04} & $30^{\circ}.7\pm 1^{\circ}.1$& $151^{\circ}.0\pm02^{\circ}.4$ & Cepheids \\
\citet{pers04} & $27^{\circ}.0\pm 6^{\circ}.0$& $127^{\circ}.0\pm10^{\circ}.0$ & Cepheids  \\
\citet{koer09} & $23^{\circ}.5\pm 0^{\circ}.4$& $154^{\circ}.6\pm01^{\circ}.2$ & Red clump stars  \\  
\citet{as10} & $23^{\circ}.0\pm 0^{\circ}.8$& $163^{\circ}.7\pm01^{\circ}.5$ & Red Clump Stars (OGLE-III data) \\
\citet{as10} & $37^{\circ}.4\pm 2^{\circ}.3$& $141^{\circ}.2\pm03^{\circ}.7$ & Red Clump Stars (MCPS data) \\
\citet{as13} & $25^{\circ}.7\pm 1^{\circ}.6$& $141^{\circ}.5\pm04^{\circ}.5$ & Red Clump Stars (MCPS Infrared data) \\
\citet{vmc12} & $26^{\circ}.2\pm 2^{\circ}.0$& $129^{\circ}.1\pm13^{\circ}.0$ & Modeling of LMC disk plane\\
\citet{haschke12} & $32^{\circ}.0\pm 4^{\circ}.0$& $114^{\circ}.0\pm13^{\circ}.0$ & OGLE-III data (RRL stars)\\
\citet{haschke12} & $32^{\circ}.0\pm 4^{\circ}.0$& $116^{\circ}.0\pm18^{\circ}.0$ & OGLE-III data (Cepheids)\\ \hline\hline
This work& $24^{\circ}.20$ &$176^{\circ}.01$ & OGLE-III data (RRab stars)\\
(Principal axes transformation method)&&& \\\hline
This work& $36^{\circ}.43$ &$149^{\circ}.08$ & OGLE-III data (RRab stars)\\
(Plane fitting procedure)&&& \\
\hline \hline
\end{tabular}
\end{table*}
\section{Metallicity gradient in the LMC}
\label{metgrad}
The presence of radial metallicity gradient in a galaxy provides clues to the 
presence of different stellar populations which might be related to its star 
formation history or to the accretion process by an external system 
\citep{bern08}.
There are some studies in the literature which hint at the presence of a 
metallicity gradient in the LMC \citep{cioni09,feas10,ata13}. On the other 
hand, some other studies claim that no such metallicity gradient exists in the 
LMC \citep{groc06,carr11,hasc_met,piat13}. These studies support the fact that 
the amount of gradient observed as a function of distance from the LMC center 
is very small compared to the error bars of the mean metallicity obtained as a 
function of distance bins from the center \citep{groc06,carr11,hasc_met,piat13}. More recently, an extensive study based 
on the metallicities obtained from the age-metallicity relation of $21$ LMC 
fields consisting of $5.5$ million stars   indicates that there exists no 
metallicity gradient in the LMC \citep{piat13}.              

\begin{figure}
\includegraphics[width=0.5\textwidth,keepaspectratio]{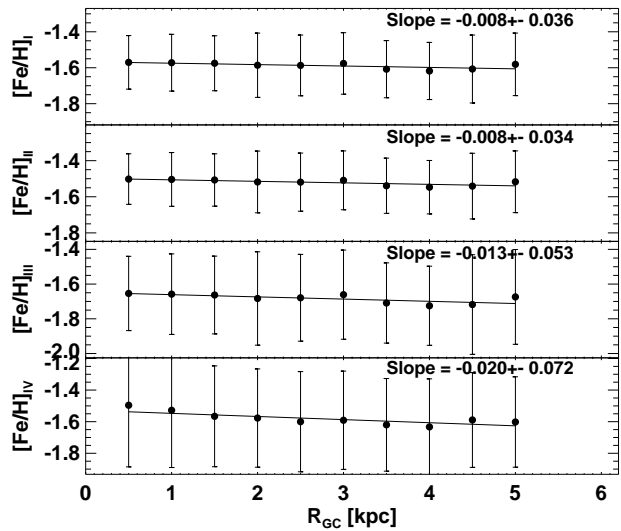}
\caption{Mean metallicity distribution of $13,095$ RRab stars as a function 
of galactocentric distance in kpc with a bin size of $0.5$ kpc. Mean 
metallicities in each bin has been obtained using the four empirical relations 
as described in the text.}
\label{lmc_grad}
\end{figure}

\begin{figure}
\includegraphics[width=0.5\textwidth,keepaspectratio]{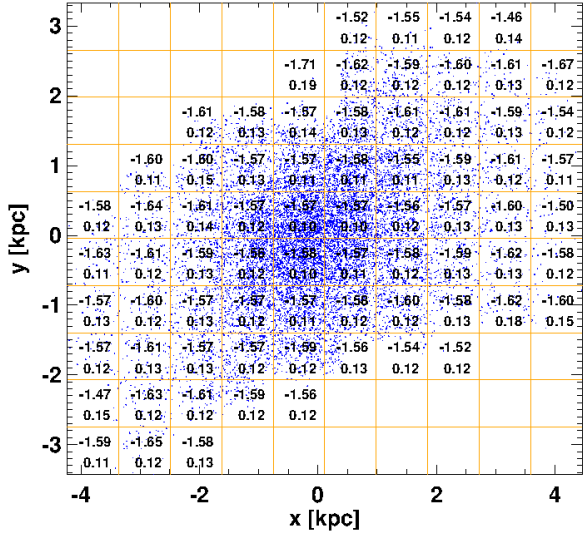}
\caption{Metallicity distribution of $13,095$ RRab stars in the LMC. 
Metallicity values are binned on a $10\times 10$ coordinate grid. In each 
bin, we compute the average metallicity and the associated errors which are 
due to the uncertainties in the determination of the Fourier parameter 
$\phi_{31}$ and are shown in each grid box.}
\label{metmap}
\end{figure}

RRL stars provide an unique opportunity to measure the metallicity of old
stellar populations and serve as a valuable means to identify the existence 
of any possible metallicity gradient in a galaxy. In order to examine the 
question of metallicity gradient in the LMC, we have resorted to the 
metallicity values of RRab stars obtained here using the four empirical 
relations as described in the section~\ref{iron} as a function of the 
galactocentric distances $(R_{GC})$. Galactocentric distances were calculated 
following the method of \citet{cioni09}, taking into consideration the viewing 
angles of the LMC disk: the inclination ($i = 24^{\circ}.20$) and the position 
angle of the line of modes $\theta_{lon} = 176^{\circ}.01$ as determined for 
RRab stars outlined above using the principal axes transformation method. 
Angular distances and position angle coordinates of all the RRab stars were 
computed using the equations given in \citet{vand101} considering the center 
of the LMC ($\alpha_{2000}=5^{\rm h}21^{\rm m} 24^{\rm s}$ and 
$\delta_{2000} = -69^{\circ}39^{\prime}$) as origin determined for the present 
sample. Fig.~\ref{lmc_grad} shows the weighted mean metallicity values 
computed using different empirical relations as a function of distance in 
bins of $0.5$ kpc from the LMC center \citep{bevi03}. 
$[Fe/H]_{I},[Fe/H]_{II},[Fe/H]_{III},
[Fe/H]_{IV}$ denote the mean metallicity values in each bin computed as a 
function of distance using Eqs.~\ref{jk96}, \ref{sm05},\ref{s04}, 
\ref{alco2000}, respectively. Least square fits to the distance and the mean 
metallicity values with their estimated errors to each of the data sets yield 
slopes of $-0.008\pm0.036$ dex~kpc$^{-1}$, $-0.008\pm0.034$ dex~kpc$^{-1}$, 
$-0.013\pm0.053$ dex~kpc$^{-1}$ and $-0.020\pm0.072$ dex~kpc$^{-1}$. 
Statistically, all these values correspond to no metallicity gradient. In the 
estimation of mean metallicity values in each distance bin, we have ensured 
that the number of stars are greater than $10$ for reliable statistics. Also, 
in the calculation of mean metallicity errors, we have taken into account the 
errors due to the uncertainties in the Fourier parameters and the systematic 
errors in each of the the empirical relations. These two errors were added 
quadratically for each star in order to estimate the mean metallicity and its 
associated error in a distance bin. All the empirical relations for metallicity 
calculations do not show any significant metallicity gradient within the 
uncertainties, consistent with the results obtained by \citet{groc06}, 
\citet{carr11}, \citet{hasc_met} and \citet{piat13}, at least in the inner 
$6$ kpc of the LMC. 
Fig.~\ref{metmap} shows the metallicity map of LMC RRab stars on a 
rectangular grid. To produce the $(x,y)-$map , the observed LMC area were 
binned on a $(10\times10)$ grid. The average values of the metallicity and 
their statistical uncertainty were estimated in each spatial bin. The 
metallicity values considered here are those of $[Fe/H]_{I}$ determined in 
this study.                
\section{Summary and Conclusions}
\label{summary}
In this paper, we have undertaken a  careful and systematic study of the RRab 
stars present in the OGLE-III catalog. The cleaned phased light curves were 
utilized  in order to obtain various Fourier parameters. All the Fourier 
parameters needed for the analysis of the RRab 
stars are provided with their errors. Precise selection criteria were employed 
to get a clean sample of the RRab stars for further analysis. The 
application of the `compatibility test' of JK96 based on the calculation of 
the deviation parameters yields $13,095$ `normal-looking' RRab stars. 
The light curves of these 13,095 RRab stars were analyzed further for determination of their metallicities and  distances enabling us to study the structure 
of the LMC.

The representation of RRab stars on $P-\phi_{31}^{V}$ diagram clearly shows the 
existence of three significant metallicity groups with mean metallicities $-1.20 \pm 0.12$ dex, 
$-1.57 \pm 0.10$ dex and $-1.89 \pm 0.09$ dex.  The corresponding absolute magnitudes of these three 
groups are obtained as $0.70\pm 0.08$ mag, $0.59 \pm 0.06$ mag and 
$0.49 \pm 0.08$ mag, respectively. Distribution of these three groups as a function of 
vertical $|z|$ distance indicates that majority of the stars belonging to each 
group are concentrated upto $\sim 10$ kpc which traces the disk of the 
LMC. The distribution beyond $|z| = 10$ kpc suggests the existence of an 
inner halo of the LMC, where most of the stars are metal poor belonging to 
groups $II$ and $III$. Since the majority of the old and metal poor stars 
($>91\%$) are located within $|z| = 10$ kpc, it may be concluded that the disk 
of the LMC has been formed much earlier than the extended halo.

The structure of the LMC has been studied using the distance distribution of 
the $13,095$ RRab stars. The coordinates $(\alpha,\delta)$ and the individual 
distances $D$ of the RRab stars were converted into the Cartesian coordinates
with the origin at the LMC center $(\alpha_{0}, \delta_{0})$ and the mean  
distance to the LMC $D_{0}$. Approximating the LMC disk as a tri-axial 
ellipsoid, we have used the principal  axes transformation of the moment of 
inertia tensor obtained from the distribution of the selected sample of    
RRab stars. The following geometrical parameters have been determined: 
inclination $i=24^{\circ}.20$ and position angle of the line of nodes 
$\theta_{lon}=176^{\circ}.01$.   

The question of existence of metallicity gradient in the LMC has been studied 
using the $P-A_{V}-[Fe/H]$ empirical relation and the $P-\phi_{31}-[Fe/H]$ 
relations. We did not find any evidence of a radial metallicity gradient in 
the LMC within the uncertainties of their values. Accurate spectroscopic 
measurements are needed to confirm findings.          
\section*{Acknowledgments}  
HPS acknowledges Indo-US Science \& Technology Forum (IUSSTF) for supporting 
the Indo-US Joint Center on Analysis of Variable Star Data. The authors 
acknowledge helpful discussions with Shashi Kanbur. SD thanks Department of 
Science \& Technology (DST), Govt. of India for support under Fast Track 
Scheme for Young Scientist in Physical Sciences.  The  study made use of 
arxiv.org/archive/astro-ph and NASA ADS databases. The authors thank the 
anonymous referee for many useful comments and suggestions that significantly 
improved the paper.            
\bibliographystyle{mn2e}
\bibliography{deb}
\end{document}